\documentclass[aps,pra,twocolumn,showpacs,superscriptaddress,8pt,longbibliography]{revtex4-2}
\usepackage{amsmath}
\usepackage{amssymb}
\usepackage[colorlinks = true,linkcolor = red,citecolor = magenta]{hyperref}
\usepackage[sort&compress]{natbib}
\usepackage{scalerel}
\usepackage[normalem]{ulem}
\usepackage{graphicx}
\usepackage{xcolor}
\usepackage{physics}
\usepackage{dsfont}
\usepackage{mathrsfs}
\usepackage{float}
\usepackage{mathtools}
\usepackage{subfigure}
\usepackage{cleveref}

\newcommand{\pcsadd}{Center for Theoretical Physics of Complex Systems, Institute for Basic Science (IBS), Daejeon 34126, Korea}
\newcommand{\ustadd}{Basic Science Program, Korea University of Science and Technology (UST), Daejeon 34113, Korea}

\newcommand{\sect}[1]{\emph{#1}.---}
\newcommand{\pts}{\ensuremath{\mathcal{PT}}}

\newcommand{\mc}{\mathcal{A}}
\newcommand{\mh}{\mathcal{H}}

\newcommand{\mf}{\mathcal{F}}

\newcommand{\mE}{\mathcal{E}}

\newcommand{\mhme}{\mh_\mE}

\newcommand{\vmE}{\vec{\mE}}
\newcommand{\vk}{\vec{k}}
\newcommand{\vkp}{\vec{\kappa}}
\newcommand{\vlp}{\vec{r}}
\newcommand{\vn}{\vec{n}}
\newcommand{\vm}{\vec{m}}

\begin{document}

\title{Anti-$\pts$ flatbands}

\author{Arindam Mallick}
  \email{marindam@ibs.re.kr}
  \affiliation{\pcsadd}

\author{Nana Chang}
  \email{nnchangqq@gmail.com}
  \affiliation{\pcsadd}
  \affiliation{Beijing National Research Center for Information Science and Technology, Tsinghua University, Beijing 100084, China}
  \affiliation{Center for Advanced Quantum Studies, Department of Physics, Beijing Normal University, Beijing 100875, People’s Republic of China}

\author{Alexei Andreanov}
  \email{aalexei@ibs.re.kr}
  \affiliation{\pcsadd}
  \affiliation{\ustadd}

\author{Sergej Flach}
  \email{sflach@ibs.re.kr}
  \affiliation{\pcsadd}
  \affiliation{\ustadd}

\date{\today}

\begin{abstract}
  We consider tight-binding single particle lattice Hamiltonians which are invariant under an antiunitary antisymmetry: the anti-\(\pts\) symmetry.
  The Hermitian Hamiltonians are defined on \(d\)-dimensional non-Bravais lattices.
  For an odd number of sublattices, the anti-\(\pts\) symmetry protects a flatband at energy \(E = 0\).
  We derive the anti-\(\pts\) constraints on the Hamiltonian and use them to generate examples of generalized kagome networks in two and three lattice dimensions.
  Furthermore, we show that the anti-\(\pts\) symmetry persists in the presence of uniform DC fields 
  and ensures the presence of flatbands in the corresponding irreducible Wannier-Stark  
  band structure.
  We provide examples of the Wannier-Stark band structure of generalized kagome networks in the presence of DC fields, and their implementation using Floquet engineering.
\end{abstract}

\maketitle

\sect{Introduction}
Flatband systems with single particle dispersionless bands in their band structure \cite{flach2014detangling,derzhko2015strongly,leykam2018artificial,leykam2018perspective,maimaiti2017compact,khomeriki2016landau, rhim2020quantum, rhim2021singular}
are important and promising platforms for exploring exotic phases and unconventional orders, due to the combined effect of macroscopic degeneracy of flatbands and applied perturbations.
Possible perturbations include disorder \cite{goda2006inverse, leykam2013flat, roy2020interplay}, nonlinear interactions \cite{leykam2013flat,danieli2021nonlinear},
and various many-body interactions \cite{kuno2020flat_qs, danieli2021quantum, heikkila2016flat, nunes2020flatband, orito2021nonthermalized}.
The presence of localized eigenstates of a flatband are argued to be useful for quantum information storage and transfer \cite{rontgen2019quantum, taie2020spatial, rontgen2020designing}, and for observing memory effects \cite{lai2016geometry}.
Remarkably, the presence of a uniform DC field leads to a Wannier-Stark (WS) ladder of \((d-1)\)-dimensional irreducible band structures in a \(d\)-dimensional lattice.
These irreducible band structures can again contain flatbands \cite{kolovsky2018topological,mallick2021wannier}.
Being fine-tuned by nature, finding flatband Hamiltonians is in general a challenging problem.
Multiple methods were developed to generate flatbands in translationally invariant systems that are based on fine-tuning \cite{maimaiti2017compact,maimaiti2019universal, maimaiti2021flatband}, line graphs \cite{mielke1991ferromagnetism},
origami rules \cite{dias2015origami}, repetition of miniarrays \cite{morales2016simple}, and application of magnetic field \cite{creutz1999end, khomeriki2016landau, moller2018synthetic, yu2020isolated}.

Flatbands can also emerge as a consequence of a symmetry.
Local and latent symmetries have been shown to generate flatbands~\cite{roentgen2018compact, morfonios2021flat}.
The other class of symmetries are global symmetries of the Hamiltonian.
A global symmetry is associated with a symmetry operator \(\Gamma\) which is either unitary or antiunitary.
A single particle Hamiltonian \(\mh\) is antisymmetric if the following relation holds: \(\Gamma \cdot \mh \cdot \Gamma^{-1} = -\mh\).
The antisymmetry implies that for each eigenvalue \(E\) with eigenvector \(\ket{\psi_E}\) there exists the negative eigenvalue \(-E\) with eigenvector \(\Gamma \ket{\psi_E}\).
If the total number of eigenvalues is odd, it follows that at least one of them is zero.
Translationally invariant lattice Hamiltonians are characterized by the number of their sublattices.
Transforming the Hamiltonian into Bloch momentum space and observing \(\Gamma(\vk) \cdot \mh(\vk) \cdot \Gamma^{-1}(\vk) = -\mh(\vk)\) results in a macroscopically degenerated symmetry-protected \(E = 0\) flatband for an odd number of sublattices.

One such example is the chiral symmetry that is realized by a unitary operator \(\Gamma\).
The chiral Hamiltonian in momentum space turns bipartite,
\(
  \mh(\vk) =
  {\scriptsize{\begin{pmatrix}
    \mathbb{O} & \mathbb{T}(\vk) \\
    \mathbb{T}^\dagger(\vk) & \mathbb{O} \\
  \end{pmatrix} }}
\), 
where \(\mathbb{O}\) is a null matrix and \(\mathbb{T}(\vk)\) is a rectangular matrix.
Chiral flatband models and exhausting flatband generators have been reported for dimension \(d = 1,2,3\)~\cite{sutherland1986localization,murpetit2014chiral,ramachandran2017chiral}.

In this Letter, we explore the other possibility when the symmetry operator \(\Gamma\) is antiunitary and analyze the effect of the applied DC field. 
Only few results are known in this case.
In \(d = 2\), Green {\it et al}.~\cite{green2010isolated} introduced a family of modified kagome lattices with three sublattices with nonzero local flux distributions which have a symmetry-protected flatband at energy \(E = 0\) despite the breaking of time-reversal symmetry.
Specific members of this modified kagome family were reported in later publications as well~\cite{koch2010time,wei2021optical}.
A specific decoration of the 2D Lieb lattice was also reported to features a symmetry-protected flatband~\cite{chen2014the}.
We note that all of the respective antiunitary operators \(\Gamma = \mathcal{A}\) consist of a spatial point reflection (inversion through a point) in lattice position space \(\mathcal{P}\),
followed by a time reversal operation \(\mathcal{T}\) (usually simply an antilinear complex conjugation operation in lattice position basis): \(\mathcal{A} = \mathcal{T} \cdot \mathcal{P}\).
Therefore, all of the above examples enjoy anti-\(\pts\) Hamiltonians.

When a commensurate uniform DC field~\cite{mallick2021wannier} is applied, the band structure of the original \(d\)-dimensional Hamiltonian is modified into a Wannier-Stark ladder of irreducible \((d-1)\)-dimensional band structures with the same number of bands~\cite{maksimov2015wannierII}.
The particular case of the 2D dice lattice with three bands resulted in a WS flatband in the presence of a DC field, which was believed to be protected by the chiral (bipartite) symmetry of the original dice lattice~\cite{kolovsky2018topological}.
However, this symmetry appears to be lost in the presence of DC fields, and the flatband existence proof in Ref.~\cite{kolovsky2018topological} does not explicitly rely on it.
However, we note that the dice lattice is also invariant under anti-\(\pts\) symmetry.
As we show below, this symmetry remains, in general, intact in the presence of a nonzero DC field.

We derive the constraints for a general Hermitian Hamiltonian \(\mh\) on a \(d\)-dimensional non-Bravais lattice to be anti-\(\pts\) symmetric:
\begin{gather}
  \label{eq:APT}
  \mathcal{A} \cdot \mh \cdot \mathcal{A}^{-1} = - \mh\;.
\end{gather}
The anti-\(\pts\) symmetry condition and an odd number of sublattices are sufficient to protect at least one flatband---both in the absence and presence of DC fields.

\sect{Definitions}
We consider a Hermitian tight-binding Hamiltonian on a \(d\)-dimensional non-Bravais lattice.
Every lattice site is labeled by its unit cell index vector \(\vn = \sum_{j = 1}^d n_j \vec{a}_j\) and sublattice index \(\nu = 1, 2, \ldots, \mu\).
The numbers \(n_j\) are integers and \(\vec{a}_j\) are \(d\)-dimensional unit cell basis vectors (and are, in general, neither orthogonal nor normalized).
Similar to the lattice vector \(\vn\), we define the sublattice vectors \(\vm_\nu = \sum_{i = 1}^d m_{\nu, i} \vec{a}_i\) which locate sublattice sites relative to a unit cell: \(-1 < m_{\nu, i} < 1\).
Consequently, we label the Hilbert space basis vectors as \(\ket{\nu, \vn}\).
The single-particle translationally invariant Hamiltonian reads
\begin{align}
    \mh = -\sum_{\vec{l}, \vn} \sum_{\nu, \sigma = 1}^{\mu} t_{\nu, \sigma}(\vec{l}) |\nu, \vn\rangle \langle\sigma, \vn + \vec{l}|\,.
    \label{eq:hamil}
\end{align}
The hopping amplitude \(t_{\nu, \sigma}(\vec{l}) = t^*_{\sigma, \nu}(-\vec{l})\) connects site \((\sigma, \vn + \vec{l})\) with site \((\nu,\vn)\).
Application of the Bloch theorem on the Hamiltonian~\eqref{eq:hamil} block diagonalizes it in the quasimomentum basis \(\{|\vk\rangle\}\). Each block is a \(\mu \times \mu\) matrix acting only on the sublattice space:
\begin{align}
    \mh(\vk) \coloneqq -\sum_{\vec{l}} \sum_{\nu, \sigma = 1}^{\mu} t_{\nu, \sigma}(\vec{l}) e^{i \vk\cdot \vec{l}} \ketbra{\nu}{\sigma}.
    \label{eq:nondc_ham_in_momen}
\end{align}
If \(\mu\) is odd, an antisymmetry of the Hamiltonian \(\mh(\vk)\) results in an zero eigenvalue.
If that antisymmetry holds for all \(\vk\), then the Hamiltonian possesses a zero-energy flatband.

The anti-$\pts$ symmetry operator reads
\begin{gather}
    \mc = \mathcal{T} \cdot \mathcal{P} = \mathcal{T} \cdot \sum_{\nu, \vn} e^{i \xi_\nu} \ketbra{f(\nu), -\vn - \vec{p}_\nu}{\nu, \vn}\;.
    \label{eq:symmetry_operator_form}
\end{gather}
The one-to-one map \(f(\nu)\) describes the swap of the sublattice indices upon lattice point inversion.
It is defined by the lattice geometry and it is its own inverse: \(f^{-1} = f\).
For instance, with three sublattices the only choices are \(f_1: 1 \mapsto 1,2 \mapsto 2,3 \mapsto 3\), and \(f_2: 1 \mapsto 1,2 \mapsto 3,3 \mapsto 2\) (up to a freedom of the sublattice index relabeling).
The 2D Lieb and kagome lattices implement \(f_1\), while the 2D dice lattice implements \(f_2\).
Inversion in position space results in inverting the sign of a unit cell vector \(\vec{n} \mapsto -\vec{n}\).
However, the inversion can map a given sublattice point of unit cell \(\vec{n}\) into one of the neighboring cell of \(-\vec{n}\).
Therefore, we had to introduce the lattice vectors \(\vec{p}_\nu\) in Eq.~(\ref{eq:symmetry_operator_form}).
\(\vec{p}_\nu\) relates the sublattice vectors: \(\vec{m}_\nu + \vec{m}_{f(\nu)} = \vec{p}_\nu\).
The gauge phases \(\xi_\nu\) relate to the magnetic flux distributions (if present) in the models of interest.
Since we consider an odd number of sublattices, it follows that \(\mc^2 = \mathds{1}\) (see  Section \ref{sec:lattice_vecconditions_antipt} of Supplemental Material for details).
This implies the following constraints: \(\vec{p}_\nu = \vec{p}_{f(\nu)}\) and \(\xi_\nu = \xi_{f(\nu)}\).
For instance, in the case of three sublattices, \(f_1\) allows for three independent gauge phases while \(f_2\) allows for only two independent gauge phases.
Combining Eqs.~\eqref{eq:APT} and~\eqref{eq:symmetry_operator_form}, we arrive at the following constraints on the hoppings for an anti-$\pts$ symmetric Hamiltonian~\eqref{eq:hamil}:
\begin{align}
    e^{-i \xi_\nu + i \xi_\sigma} t^*_{\nu, \sigma}(\vec{l}) =
    - t_{f(\nu), f(\sigma)}(- \vec{l} + \vec{p}_\nu - \vec{p}_\sigma)\;.
    \label{eq:hooping_main_restrict}
\end{align}
The above constraint on the hoppings can be used to efficiently construct anti-\(\pts\) symmetric Hamiltonians.
For a single sublattice (e.g., a Bravais lattice) the above condition~\eqref{eq:hooping_main_restrict} reduces to \(t^* (\vec{l}) = - t(- \vec{l})\).
At the same time, the Hermiticity of the Hamiltonian enforces \(t^* (\vec{l}) = t(-\vec{l})\).
Both conditions can only be satisfied for the trivial case of no hopping \(t(\vec{l}) = 0\).
Therefore, the anti-\(\pts\) symmetry requires two or more sublattices.

\sect{Anti-\(\pts\) protected flatbands} Let us project both sides of Eq.~\eqref{eq:APT} onto the \(\vk\)-space:
\begin{gather}
  \label{eq:symm_for_each_k_nodc}
  \mc(\vk) \cdot \mh(\vk) \cdot \mc{(\vk)}^{-1} = - \mh(\vk).
\end{gather}
For a Hamiltonian satisfying Eq.~\eqref{eq:hooping_main_restrict}, the anti-\(\pts\) operator~\eqref{eq:symmetry_operator_form} transforms as
\begin{align}
  \mc(\vk) = \mathcal{T}_s \cdot \sum_{\nu = 1}^\mu e^{i \xi_\nu} e^{-i \vk \cdot \vec{p}_\nu} |f(\nu)\rangle\langle \nu|,
  \label{eq:APTk}
\end{align}
where \(\mathcal{T}_s\) is a complex conjugation operator and it acts only on the sublattice space.
For an odd number of sublattices \(\mu\), one of the \(\mu\) eigenvalues of \(\mh(\vk)\) is zero.
As this is true for all \(\vk\), it follows that one of the bands must be flat with energy equal to zero.

\begin{figure}
  \subfigure[]{\includegraphics[width = 0.8\columnwidth]{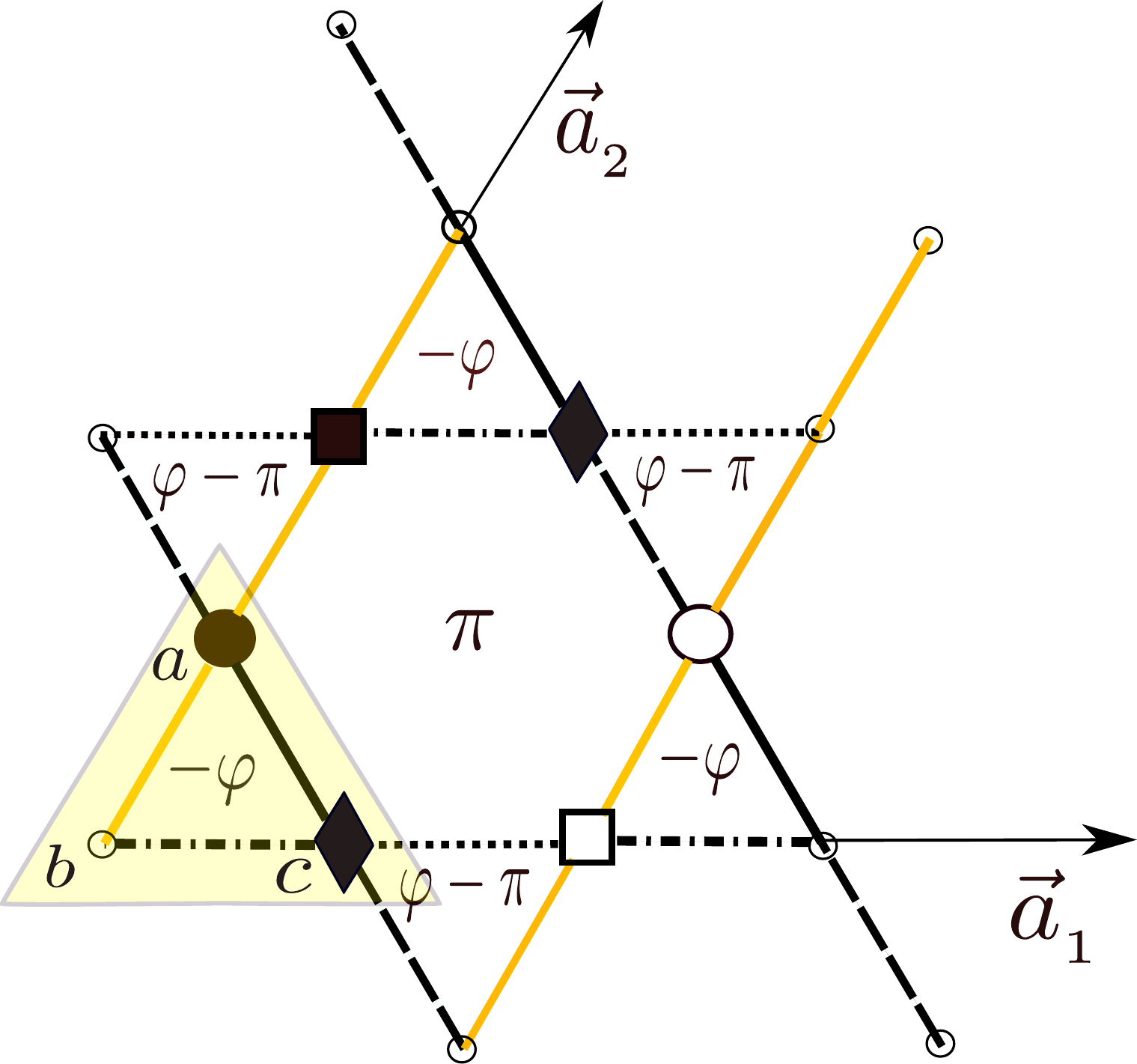}
  \label{fig:kagome_schematic}}
  \subfigure[]{\includegraphics[width = 0.4\columnwidth]{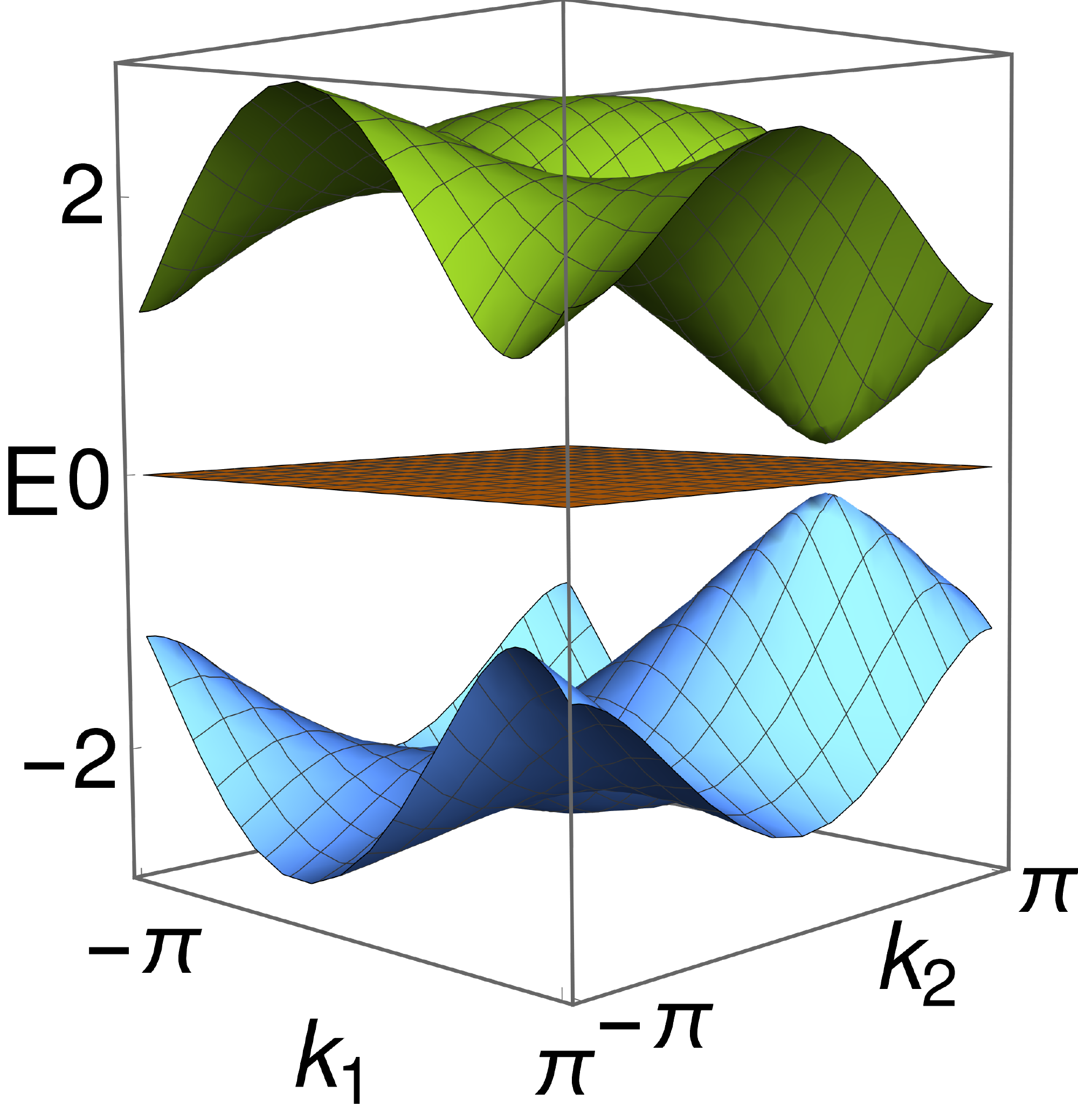}
  \label{fig:kagome_spec_flux_piby5_015_spectra}}
  \subfigure[]{\includegraphics[width = 0.39\columnwidth]{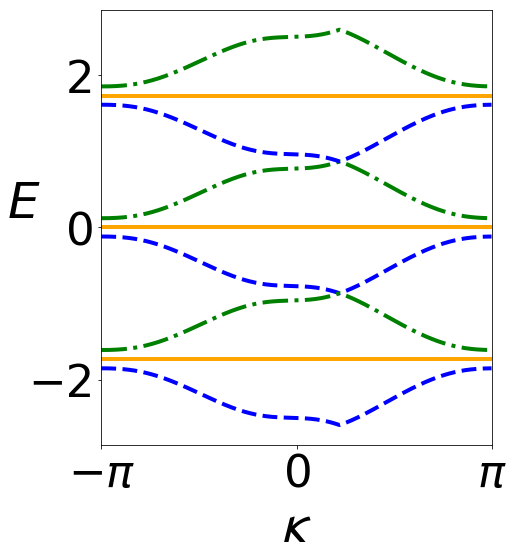}
  \label{fig:kagome_spectra_dc}}
  \caption{(a) The anti-\(\pts\) symmetric generalized 2D kagome lattice.
    The lattice sites are shown by small empty black circles.
    A single unit cell is shown within a shaded triangle with the sublattice sites \(a\) (\(\nu = 1\)), \(b\) (\(\nu = 2\)), and \(c\) (\(\nu = 3\)).
    The hoppings \(t_{3,1}(1,-1) = -1\) (black dashed lines), 
    \(t_{2,3}(0,0) = e^{i \varphi}\) (black dashed-dotted lines),
    \(t_{2,3}(-1,0) = e^{-i\varphi}\) (black dotted lines), \(t_{1,2}(0,1) = t_{1,2}(0,0) = t\) (yellow solid lines), \(t_{1,3}(0,0) = 1\) (solid black lines).
    The fluxes \(\varphi\) induced by anti-\(\pts\) symmetric complex hopping choices are denoted inside each plaquette, with all fluxes computed counter-clockwise.
    The compact localized eigenstate at the anti-\(\pts\) flatband energy \(E = 0\) has nonzero wave-function amplitudes indicated by large circles, diamonds, and squares (for more details, we refer to the main text).
    (b) Band structure \(E(k_1,k_2)\) for \(\varphi = \frac{\pi}{5}\) and \(t = 0.15\).
    (c) Three subsequent irreducible Wannier-Stark band structures computed using Eq.~\eqref{eq:gen_main_eigen} for the DC field direction \(\vec{a}_1 + \vec{a}_2\).
    The field strength \(|\vec{\mE}| = 2\).
  }
  \label{fig:kagome}
\end{figure}

In Fig.~\ref{fig:kagome}, we show an anti-$\pts$ symmetric generalized 2D kagome lattice with an \(E = 0\) flatband compatible with Eq.~\eqref{eq:hooping_main_restrict}.
The sublattice vectors are \(\vm_1 = \frac{1}{2} \vec{a}_2\), \(\vm_2 = \vec{0}\), and \(\vm_3 = \frac{1}{2} \vec{a}_1\), while \(f(\nu) = \nu\), \(\vec{p}_1 = \vec{a}_2\), \(\vec{p}_2 = \vec{0}\), and \(\vec{p}_3 = \vec{a}_1\).
The hopping parameters are detailed in the caption of Fig.~\ref{fig:kagome}.
Diagonalizing the Hamiltonian \(\mh(\vk)\) for this choice of parameters, we obtain three bands 
(see Section \ref{app:kagome_cond} of Supplemental Material).

The anti-\(\pts\) band structure is shown in Fig.~\ref{fig:kagome_spec_flux_piby5_015_spectra}.
The anti-\(\pts\) flatband supports eigenstates which are compact localized states (CLSs) occupying three unit cells as shown in Fig.~\ref{fig:kagome_schematic}.
The CLS amplitudes up to normalization are \(\equiv -t\) (black diamonds), \(\equiv e^{i \varphi}\) (black filled circle), \(\equiv e^{-i \varphi}\) (empty big circle),
\(\equiv +1\) (black filled square), and \(\equiv -1\) (empty square).

\begin{figure}
  \subfigure[]{\includegraphics[width = 0.3\columnwidth]{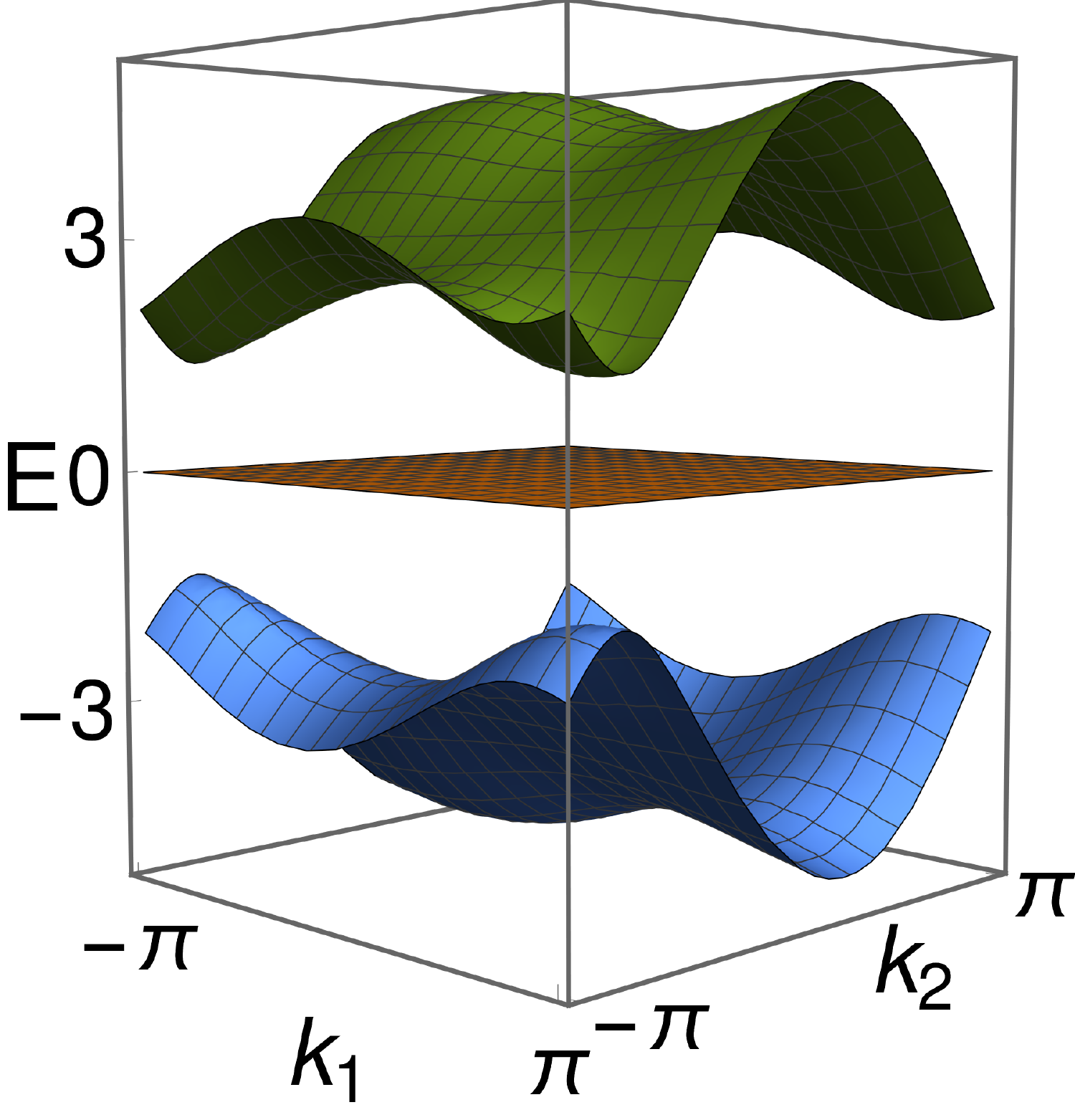}
  \label{fig:kagome3d_flux_piby5_k3_piby7_t_015_spectra}}
  \subfigure[]{\includegraphics[width = 0.3\columnwidth]{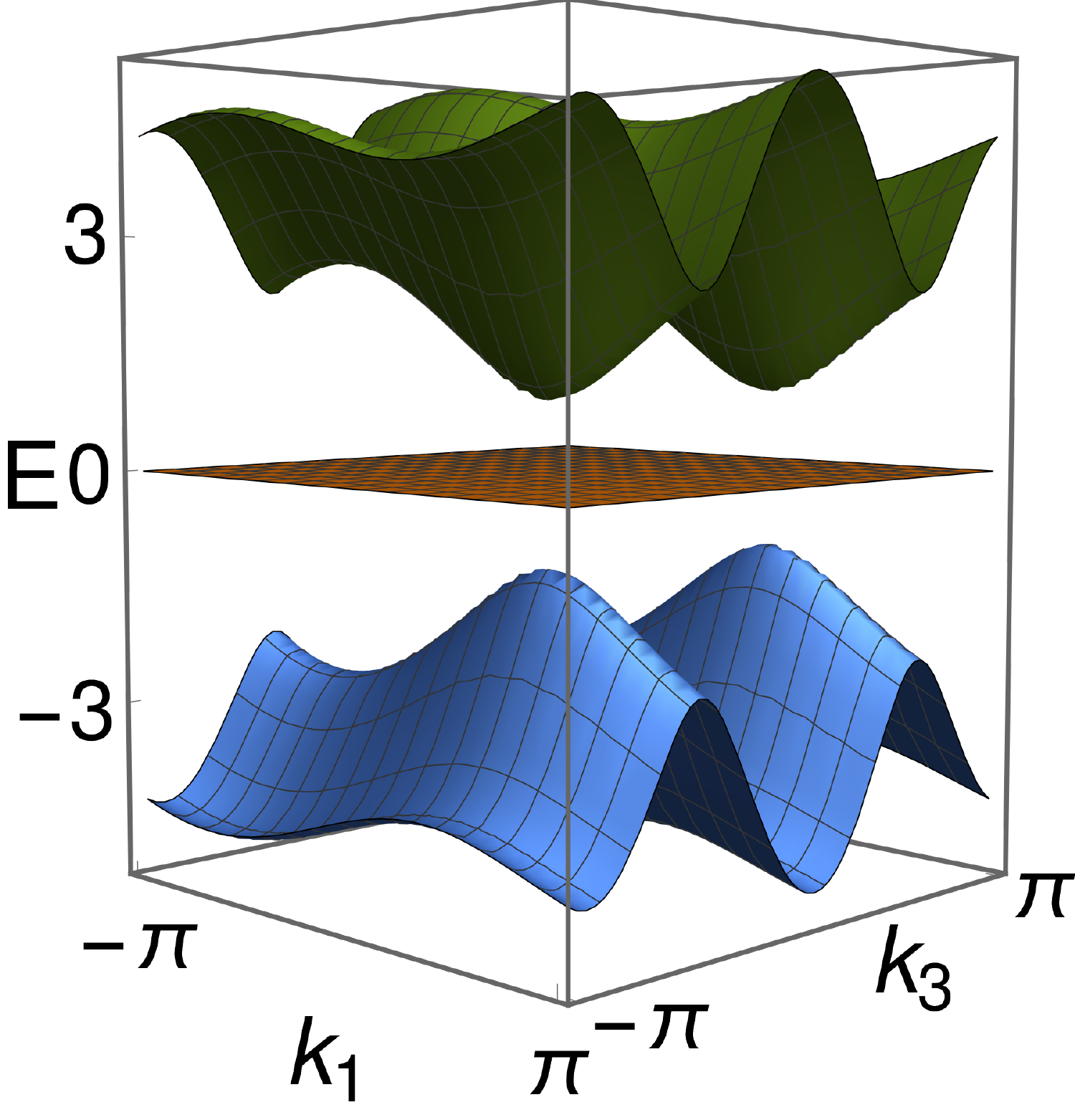}
  \label{fig:kagome3d_flux_piby5_k2_piby7_t_015_spectra}}
  \subfigure[]{\includegraphics[width = 0.3\columnwidth]{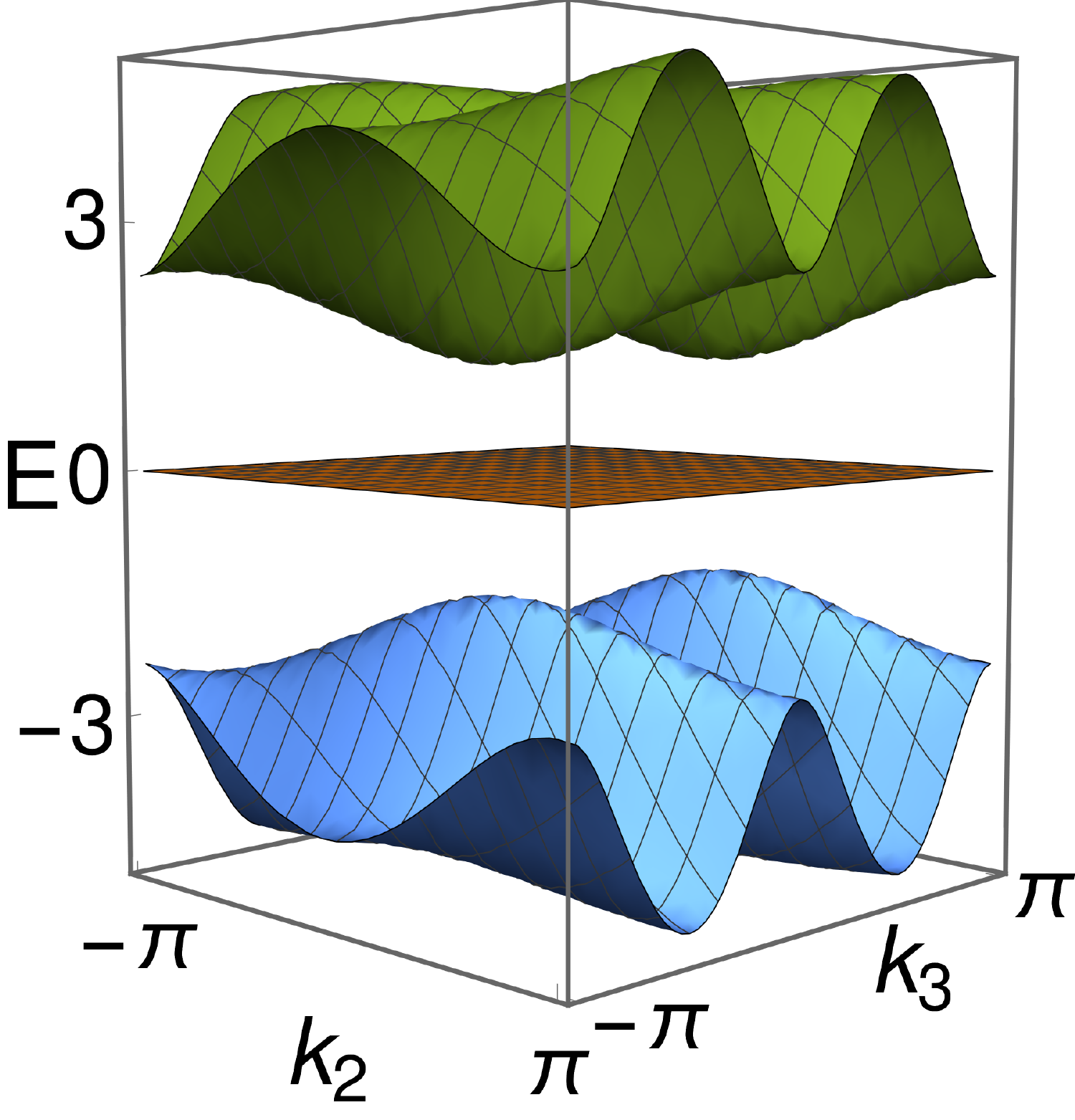}
  \label{fig:kagome3d_flux_piby5_k1_piby7_t_015_spectra}}
  \subfigure[]{\includegraphics[width = 0.45\columnwidth]{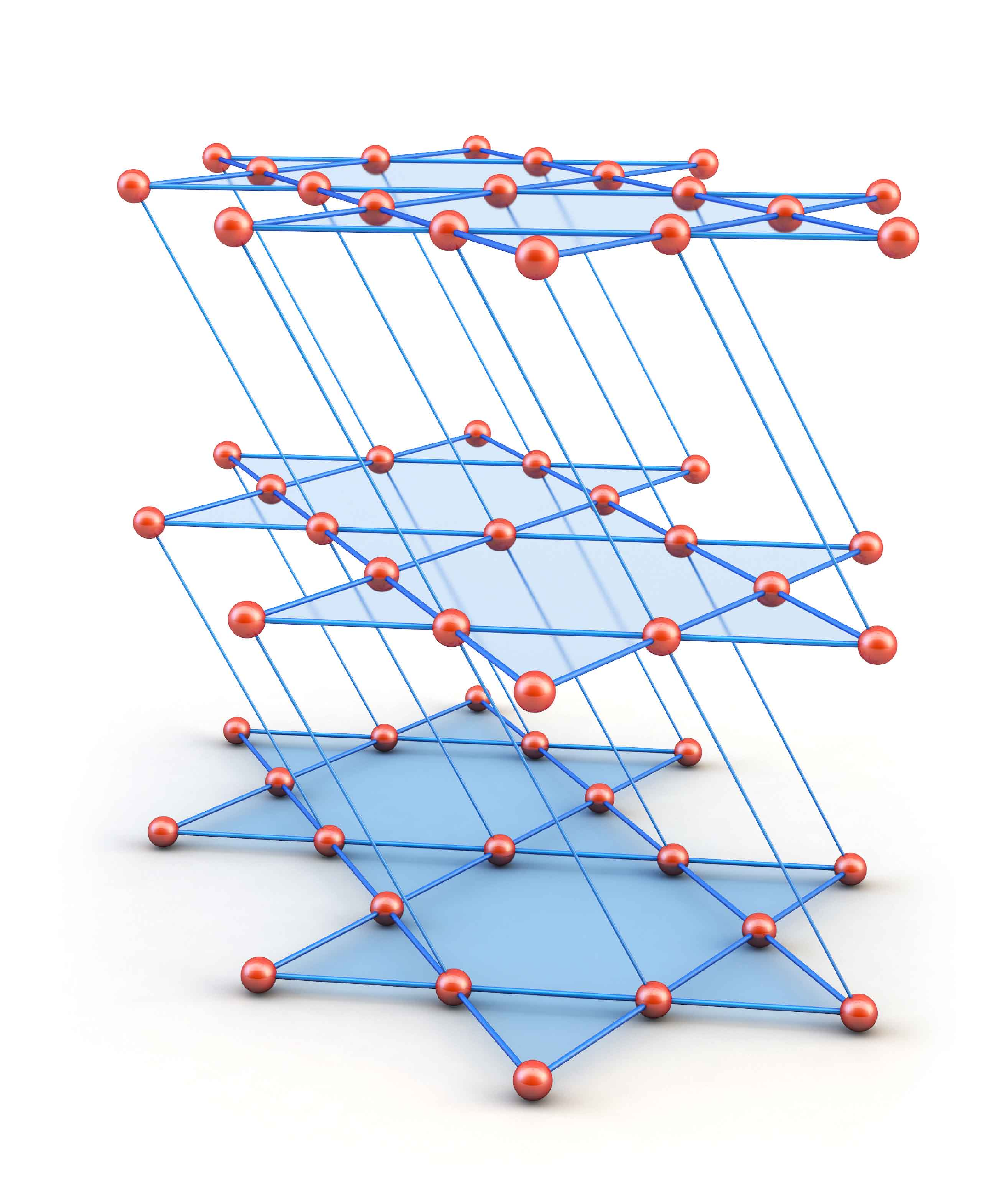}
  \label{fig:kagome3d_schematic}}
  \subfigure[]{\includegraphics[width = 0.45\columnwidth]{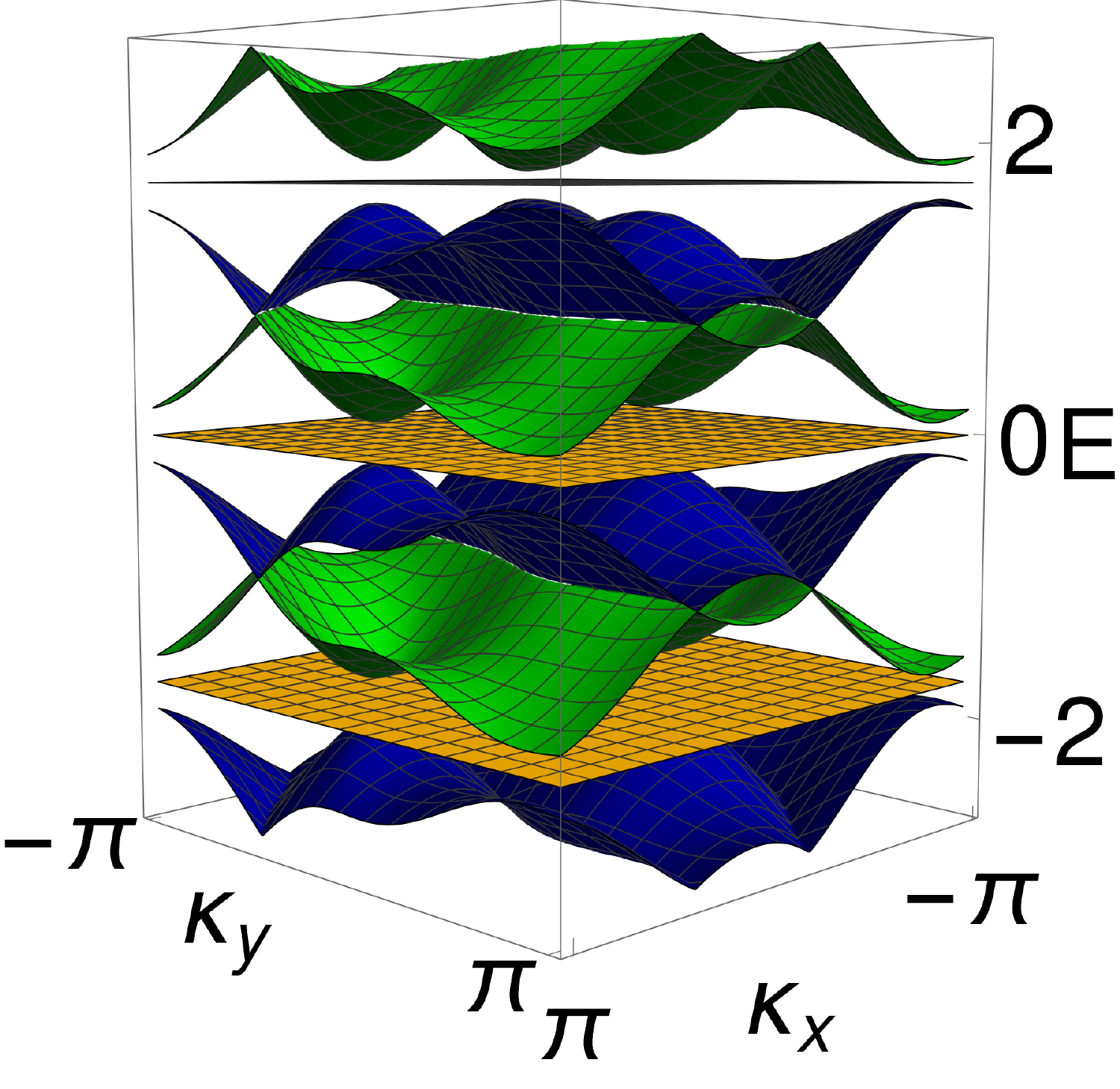}
  \label{fig:3dkagome_spectra_dc}}
  \caption{The anti-\(\pts\) symmetric generalized 3D kagome lattice.
    (a)-(c) Three constrained band structures:
    (a) \(E(k_1, k_2, k_3 = \frac{\pi}{7})\), (b) \(E(k_1, k_2 = \frac{\pi}{7}, k_3)\), (c) \(E(k_1 = \frac{\pi}{7}, k_2, k_3)\).
    (d) The lattice structure. The sites are denoted by small solid red spheres.
    The hopping connections within each 2D kagome plane are the same as in Fig.~\ref{fig:kagome_spec_flux_piby5_015_spectra}.
    The intraplane hoppings \(t_{1,2}(0,0,1) = 2\) and \(t_{1,2}(0,1,-1) = 2\).
    (e) Three subsequent irreducible Wannier-Stark band structures \(E_{\gamma, a}(\kappa_x, \kappa_y)\) computed using Eq.~\eqref{eq:gen_main_eigen}
    for the field direction \((2,2,3) \equiv 2\vec{a}_1 + 2 \vec{a}_2 + 3\vec{a}_3\).
    The field strength \(|\vec{\mE}| = \sqrt{7}\).}
  \label{fig:3dkagome}
\end{figure}

To arrive at a 3D version of the kagome lattice, shown in Fig.~\ref{fig:kagome3d_schematic},
we stack the 2D kagome lattices shown in Fig.~\ref{fig:kagome_schematic} on top of each other vertically with \(|\vec{a}_3| = 1\).
Two additional hoppings connect neighboring 2D kagome planes: \(t_{1,2}(0,0,1) = 2\) and \(t_{1,2}(0,1,-1) = 2\).
The spectrum is now a function of three reciprocal momenta \((k_1, k_2, k_3)\).
In Fig.~\ref{fig:kagome3d_flux_piby5_k3_piby7_t_015_spectra}-\ref{fig:kagome3d_flux_piby5_k1_piby7_t_015_spectra}, we plot three different 3D intersections of the band structure \(E(k_1, k_2, k_3)\).
All of them contain an anti-\(\pts\) flatband at zero energy.

\sect{Anti-\(\pts\) protected Wannier-Stark flatbands} 
We now outline and prove the survival of the anti-\(\pts\) symmetry in the presence of a uniform DC field \(\vmE\) for an anti-\(\pts\) symmetric Hamiltonian.
The DC field adds an on-site potential term in the Hamiltonian~\eqref{eq:hamil} and the full Hamiltonian reads
\begin{align}\label{eq:HDC}
  \mhme = \vmE \cdot \hat{r} + \mh \;.
\end{align}
Here we defined the lattice position operator as \(\hat{r} = \sum_{\nu, \vn} (\vn + \vm_\nu) \ket{\nu, \vn}\bra{\nu, \vn}\).
The DC field term \(\vmE \cdot \hat{r}\) changes sign under the application of the anti-\(\pts\) operator \(\mc\) due to lattice reflection \(\mathcal{P}\): \((\vn + \vm_\nu) \mapsto -(\vn + \vm_\nu)\).
Together with condition~\eqref{eq:hooping_main_restrict}, this ensures
\begin{align}
    \mc \cdot \mhme \cdot \mc^{-1} = -\mhme.
    \label{eq:ham-dc-anti-pt}
\end{align}

The application of the uniform DC field breaks translation invariance and eliminates the band structure for generic directions of the DC field.
However, for special field directions, translation invariance is broken only partially and a WS band structure emerges as translation invariance is preserved in the direction orthogonal to the field.
We refer to such field directions as \emph{commensurate}~\cite{mallick2021wannier}.
The unit cell and sublattice coordinates along the field, \(z\) and \(z_\nu\), respectively, are defined as \(z = \frac{1}{\mf}\vmE \cdot \vn, z_\nu = \frac{1}{\mf}\vmE \cdot \vm_\nu\) with the scaling factor \(\mf\) ensuring that \(z\) taking integer values.
The directions perpendicular to \(\vmE\) are parametrized by a \(d-1\) dimensional integer vector \(\vec{\eta}\) (see Section \ref{sec:z_w_parametrization} of Supplemental Material for details).
The Hamiltonian \(\mhme\) is translationally invariant in \(\vec{\eta}\).
With the use of the Bloch basis for \(\vec{\eta}\),
\begin{align}
  \ket{\psi_E(\vkp)} = {(2\pi)}^{\frac{1-d}{2}}\sum_{z, \nu, \vec{\eta}} \psi_E(\nu, z, \vkp)
  e^{i \vkp \cdot \vec{\eta}} \ket{\nu, z,\vec{\eta}},
\end{align}
the Hamiltonian~\eqref{eq:HDC} becomes block diagonal:
\begin{align}
  \mhme = \int\limits_{\vkp}\mhme(\vkp)d\vkp, \quad \mhme(\vkp)\ket{\psi_E(\vkp)} = E(\vkp) \ket{\psi_E(\vkp)}.
  \label{eq:dc_eigenvalue_eq}
\end{align}
Each block is infinite dimensional due to the coupling along the \(z\)-direction.
The commensurability condition for the DC field implies the persistence of a generalized translational invariance along the field direction, which goes along with an overall shift of the eigenenergies.
Following Refs.~\cite{kolovsky2018topological, mallick2021wannier}, we Fourier transform from \(z\)-space to its conjugate momentum \(q\)-space
\(
  \vec{g}_E(q, \vkp) = {(2 \pi)}^{-1/2} e^{\frac{-i E q}{\mf}} \sum_{z, \nu} e^{iq(z + z_\nu)} \psi_E(\nu, z, \vkp) \ket{\nu}
\)
to arrive at \(\mu\) coupled differential equations (see Sections \ref{app:Hamil_in_dc_field} and \ref{sec:effective_WS_hamil} of Supplemental Material for derivation):
\begin{align}
  i\frac{\partial}{\partial q} \vec{g}_E(q, \vkp) = \mhme(q, \vkp) \cdot \vec{g}_E(q, \vkp)\;.
  \label{eq:gen_evl_main}
\end{align}
The resulting Hermitian Hamiltonian
\begin{align}
  \mhme(q, \vkp) = -\frac{1}{\mf} \sum_{\vec{l}, \nu, \sigma} t_{\nu, \sigma}(\vec{l}) \ketbra{\nu}{\sigma} e^{i q (z_\nu - z_\sigma)}
  e^{i\vkp \cdot \vec{\epsilon}(\vec{l}) - \frac{i q \vmE \cdot \vec{l}}{\mf}}
  \label{eq:effec_ham_gen1}
\end{align}
is a \(\mu \times \mu\) matrix which acts on the sublattice space \(\{\ket{\nu}\}\) only, \(\vec{\epsilon}(\vec{l})\) is the hopping perpendicular to the field.
Equation~\eqref{eq:gen_evl_main} describes a unitary evolution of \(\vec{g}_E(q, \vkp)\) in \(q\)-space,
\begin{align}
  \vec{g}_E(q, \vkp) = U(q, \vkp) \cdot \vec{g}_E(0, \vkp),
  \label{eq:gen_unit_ev}
\end{align}
where \(U(q, \vkp)\) is a \(q\)-ordered exponential of the integrated \(\mhme(q, \vkp)\):
\begin{align}
  & U(q, \vkp) = \mathds{1} + (- i) \int_{q' = 0}^q dq' \mhme(q', \vkp) \notag \\
  & + {(-i)}^2 \int_{q' = 0}^q \int_{q'' = 0}^{q'} dq' dq'' \mhme(q', \vkp) \mhme(q'', \vkp) + \cdots\;.
  \label{eq:q_expansion_unitary}
\end{align}

By construction, \(\vec{g}_E(2\pi, \vkp) = e^{-\frac{2 \pi i E}{\mf}} \Lambda(2\pi) \cdot \vec{g}_E(0, \vkp)\), where the matrix \(\Lambda(q)\) is diagonal with entries \(\Lambda_{\nu\nu}(q) = e^{i q z_\nu}\).
Then, from Eq.~\eqref{eq:gen_unit_ev} and the above periodicity condition, we arrive at the eigenvalue problem on the WS bands:
\begin{align}
  \big[\Lambda^\dagger(2\pi) \cdot U(2\pi, \vkp)\big] \cdot \vec{g}_E(0, \vkp) = e^{-\frac{2 \pi i E}{\mf}} \vec{g}_E(0, \vkp)\;.
  \label{eq:gen_main_eigen}
\end{align}
The spectrum of \(\mhme\) is obtained by solving the above eigenproblem,
\begin{align}\label{eq:eig_val}
  E \equiv E_{\gamma, a}(\vkp) = \mf a + \frac{i \mf}{2\pi} \ln\left[\lambda_{\gamma}(\vkp)\right],
\end{align}
where \(a \in \mathbb{Z}\) and \(\lambda_\gamma\) are the eigenvalues of the \(\mu \times \mu\) unitary matrix \(\Lambda^\dagger(2\pi) \cdot U(2\pi, \vkp)\).
The irreducible WS band structure is obtained by choosing a particular value of $a$, e.g., \(a = 0\).
The entire spectrum is generated by a parallel shift of the irreducible band structure and is parametrized by the band indices \((\gamma, a)\).

We now arrive at the formulation of our anti-\(\pts\) \textit{theorem} in the presence of the commensurate DC field:
\textit{If the original Hamiltonian \(\mh\) has an odd number of sublattices and is anti-\(\pts\) symmetric, the irreducible WS band structure of \(\mhme\) contains at least one flatband.}

\textit{Proof:}
Indeed, the anti-\(\pts\) condition~\eqref{eq:ham-dc-anti-pt} translates into a similar condition for the effective Hamiltonian \(\mhme(q, \vkp)\),
\begin{align}
  \mhme^* (q, \vkp) = - \mathcal{M}^\dagger(\vkp) \cdot \mhme(q, \vkp) \cdot \mathcal{M}(\vkp),
  \label{eq:ham-eff-anti-pt}
\end{align}
where the \(\mu \times \mu\) unitary matrix
\begin{align}
  \mathcal{M}(\vkp) = \sum_{\nu} e^{-i \xi_\nu}e^{i \vkp \cdot \vec{\epsilon} (\vec{p}_\nu)} \ketbra{f(\nu)}{\nu}.
\end{align}
\(\vec{\epsilon} (\vec{p}_\nu)\) is the same vector function of \(\vec{l}\) as in Eq.~\eqref{eq:effec_ham_gen1} but its argument is replaced by \(\vec{p}_\nu\).
Then, from Eq.~\eqref{eq:q_expansion_unitary} it is straightforward to establish that
\begin{align}
  U^* (q, \vkp) = \mathcal{M}^\dagger(\vkp) \cdot U(q, \vkp) \cdot \mathcal{M}(\vkp).
  \label{con:dc_unita_evol}
\end{align}
We note that by definition of the commensurate DC field direction, the projection of \(\vec{p}_\nu\) along the field direction will be an integer and hence \((z_\nu + z_{f(\nu)})\) will be an integer as well (see Section \ref{app:anti_pt_under_dc_field_proof} of Supplemental Material for details).
Therefore, \(e^{2 \pi i(z_\nu + z_{f(\nu)})} = 1\).
Since the operator \(\mathcal{M}(\vkp)\) maps the sublattice vector \(\ket{\nu}\) to \(\ket{f(\nu)}\), it follows that 
\begin{align}
  \Lambda^\dagger(2\pi) = \mathcal{M}(\vkp) \cdot \Lambda(2\pi) \cdot \mathcal{M}^{\dagger}(\vkp)\;.
  \label{con:lambda_dc}
\end{align}
We use the relations~\eqref{con:dc_unita_evol} and~\eqref{con:lambda_dc} to rewrite the eigenvalue problem~\eqref{eq:gen_main_eigen} into the following form (see Section \ref{app:anti_pt_under_dc_field_proof} of Supplemental Material):
\begin{align}
  \left[\Lambda^\dagger(2\pi) \cdot U(2\pi, \vkp) \right] \cdot \big[\mathcal{M}(\vkp)\cdot \vec{g}^*_E(0, \vkp)\big] \notag \\
  = e^{\frac{2 \pi i E}{\mf}} \big[\mathcal{M}(\vkp)\cdot \vec{g}^*_E(0, \vkp)\big]\;.
  \label{eq:gen_main2_eigen}
\end{align}
Equations~\eqref{eq:gen_main_eigen} and~\eqref{eq:gen_main2_eigen} imply that the eigenvalues of the unitary operator \([\Lambda^\dagger(2\pi) \cdot U(2\pi, \vkp)]\) come in pairs \((e^{-\frac{2 \pi i E(\vkp)}{\mf}}, e^{\frac{2 \pi i E(\vkp)}{\mf}})\).
For an odd number of sublattices \(\mu\), the number of eigenvalues of the operator \([\Lambda^\dagger(2\pi) \cdot U(2\pi, \vkp)]\) is also odd.
Therefore, at least one eigenvalue satisfies \(e^{-\frac{2 \pi i E(\vkp)}{\mf}} = e^{\frac{2 \pi i E(\vkp)}{\mf}}\) with \(E(\vkp)\) being \(\vkp\) independent and a multiple of \(\frac{\mf}{2}\).
Therefore, the irreducible WS band structure contains at least one anti-\(\pts\) symmetry protected flatband. \(\square\)

We check the validity of the above theorem by computing WS band structures with~\eqref{eq:gen_main_eigen} for the 2D kagome lattice in Fig.~\ref{fig:kagome} and 3D kagome lattice in Fig.~\ref{fig:3dkagome}.
Details on the field direction and strength are provided in the corresponding captions.
We observe and confirm the presence of anti-$\pts$ protected WS flatbands in Fig.~\ref{fig:kagome_spectra_dc} for the 2D case and in Fig.~\ref{fig:3dkagome_spectra_dc} for the 3D case.

\sect{Experimental realizations}
Flatband models have already been designed in metallic systems~\cite{kajiwara2016observation}, photonic lattices~\cite{zong2016observation, mukherjee2017observation, xia2016demonstration, xia2020observation, yan2020flatband, tang2020photonics}, and ultracold atoms in optical lattices \cite{jo2012ultracold}.
The unperturbed kagome lattices introduced above can be tested in similar setups~\cite{zong2016observation, jo2012ultracold} by proper design of hopping parameters with artificial gauge fields.

To observe the Wannier-Stark effect in optical lattices with ultracold atomic gases, they can be tilted, so the gravitational field acts as a DC field source~\cite{anderson1998macroscopic}.
Moreover, one can study the impact of an electric DC field on centrosymmetric lattices as reported in a very recent experiment on diamond~\cite{desantis2021investigation}.
Another option for implementing the WS Hamiltonians is to use Floquet engineering following recent experiments, which implemented Floquet Hamiltonians using ultracold atoms~\cite{bordia2017periodically, eckardt2017colloquium}.
The spectrum of WS Hamiltonians can be mapped onto that of periodically driven systems~\cite{oka2019floquet} and vise versa. 
Mapping the frequency-space of a Floquet \((d-1)\)-dimensional lattice Hamiltonian to a new spatial dimension produces an effective static Hamiltonian in a \(d\)-dimensional lattice with WS potential.
In this case, the single set of Floquet bands, which are periodic in energy, unfolds into an infinitely repeated tower of Wannier-Stark bands.
The details of the mapping between our WS Hamiltonians in 2D (3D) kagome networks and the Hamiltonians having Floquet Peierls phases in 1D (2D) diamond lattices are provided in Section \ref{supp:Floquet_mapp} of the Supplemental Material.

\sect{Discussion and conclusions}
We considered tight-binding lattice Hamiltonians on \(d\)-dimensional non-Bravais lattices, which are invariant under the anti-\(\pts\) symmetry.
We proved that the anti-\(\pts\) symmetry protects a flatband at energy \(E = 0\) for odd numbers of sublattices.
We derived the precise anti-\(\pts\) constraints on the Hamiltonian and used them to generate examples of generalized kagome networks.
Remarkably the anti-\(\pts\) symmetry persists in the presence of uniform DC fields.
We prove that the corresponding irreducible Wannier-Stark band structures will again contain anti-\(\pts\) protected flatbands.
We demonstrate the validity of our results by computing examples of the Wannier-Stark band structure of generalized 2D and 3D kagome networks in the presence of DC fields.

The zero-energy flatbands reported in Refs.~\cite{green2010isolated, koch2010time,wei2021optical, chen2014the} belong to the anti-\(\pts\) class.
They were reported for specific choices of hoppings for two-dimensional lattices.
Our results also explain the persistence of the flatband in the dice lattice~\cite{kolovsky2018topological} in the presence of the DC field.
The original proof relied on specific properties of the hopping network, and subsequent conjectures attempted to connect the proof to the bipartiteness of the unbiased lattice.
Actually, the unbiased dice lattice is both chiral and anti-\(\pts\) symmetric.
Therefore, its \(E = 0\) flatband is protected by both the chiral and the anti-\(\pts\) symmetries.
Adding a DC field destroys the chiral symmetry but preserves the anti-\(\pts\) symmetry.
Therefore, the emerging WS flatbands in the irreducible WS band structure are protected by the anti-\(\pts\) symmetry.
Anti-\(\pts\) networks do not need to be bipartite and our proof is valid for any \(d\)-dimension with arbitrary number of sublattices.

Our study focused on spinless single particle translationally invariant Hermitian Hamiltonians on non-Bravais lattices.
Our results also apply to a particle with an integer spin (or other internal degrees of freedom, e.g., orbital degrees of freedom) including spin-orbit coupling on a Bravais lattice.
The impact of disorder, many-body interactions, nonlinearities, or non-Hermiticity on our system are possible interesting directions for future investigations.
We expect that methods developed to analyze the impact of these perturbations for other flatband models might be helpful in our setting as well.
It is also interesting to study the case of incommensurate DC field directions that are expected to generate quasicrystalline structures.

\sect{Acknowledgments} This work was supported by Institute for Basic Science in Korea (No.~IBS-R024-D1). N.C.~acknowledges financial support from the China Scholarship Council (No.~CSC-201906040021).
We thank Jung-Wan Ryu for helpful discussions.


\widetext

\section*{Supplemental Material: Anti-\(\pts\) flatbands}

\subsection{Anti-\(\pts\) symmetry conditions for (sub)lattice vectors}
\label{sec:lattice_vec_antipt}

We label every lattice site \(\vlp\) by its unit cell index vector \(\vn = \sum_{i = 1}^d n_i \vec{a}_i\) and a sublattice index \(\nu = 1, 2, \ldots \mu\) where \(\mu\) is the number of sublattices.
Here \(\vec{a}_i\) is the lattice basis vector for unit cell, \(\mu\) is the number of sublattices, and \(n_i\) always take integer values.
Note that the vectors \(\vec{a}_1, \vec{a}_2,\dots,\vec{a}_d\) are complete and linearly independent, but they need not to be orthogonal nor normal.
We can express sublattice positions measured with respect to a unit cell vector \(\vn\), in terms of the basis vectors: \(\vm_\nu = \sum_{i = 1}^d m_{\nu, i} \vec{a}_i\) with \(|m_{\nu, i}| < 1\) for all \(\nu, i\).
The combination \(\nu, \vn\) uniquely identifies the lattice vector \(\vlp = \vm_\nu + \vn \equiv (\nu, \vn)\).
Therefore we can use them to label the Hilbert space basis vectors as \(\ket{\nu, \vn}\).
The lattice shift vector \(\vec{p}_\nu\) can also be written as: \(\vec{p}_\nu = \sum_{i = 1}^d p_{\nu,i} \vec{a}_i\) where \(p_{\nu,i}\) are integers.

The reflection of the position operator
\begin{gather*}
    \hat{r} = \sum_{\vn, \nu} (\vm_\nu + \vn) \ketbra{\nu, \vn}{\nu, \vn}
\end{gather*}
under the antisymmetry operation \(\mc\) reads
\begin{align}
    & \mc \cdot \hat{r} \cdot \mc^{-1} = -\hat{r}~; \notag \\
    & \Rightarrow \sum_{\vn, \nu} (\vm_\nu + \vn) \ketbra{f(\nu), -\vn - \vec{p}_\nu}{f(\nu), -\vn - \vec{p}_\nu}
    = -\sum_{\vn, \nu} (\vm_{f(\nu)} + \vn) \ketbra{f(\nu), \vn}{f(\nu), \vn} \notag \\
    & \Rightarrow \sum_{\vn, \nu} (\vm_\nu - \vn - \vec{p}_\nu) \ketbra{f(\nu), \vn}{f(\nu),\vn}
    = -\sum_{\vn, \nu} (\vm_{f(\nu)} + \vn) \ketbra{f(\nu), \vn}{f(\nu), \vn}
    \Rightarrow \vm_\nu - \vec{p}_\nu = - \vm_{f(\nu)}\;.
    \label{cond:antipt_lattice_pos}
\end{align}
By definition the sublattice vectors satisfy \(0 \leq |m_{\nu, i}| < 1, 0 \leq |m_{f(\nu), i}| < 1\) and \(p_i(\nu)\) is always integer.
Therefore the components of the shift vector \(\vec{p}_\nu\) can only take values from \(\{0, \pm 1\}\) for any \(\nu\).

\subsection{Implications of \(\mc^2 \propto \mathds{1}\)}
\label{sec:lattice_vecconditions_antipt}

The operator \(\mc\) consists of a spatial reflection and a complex conjugation operation in position basis:
\begin{gather}
    \mc = \mathcal{T} \cdot \mathcal{P} = \mathcal{T} \cdot \sum_{\nu, \vn} e^{i \xi_\nu} \ketbra{f(\nu), -\vn - \vec{p}_\nu}{\nu, \vn}.
    \label{eq:symmetry_operator_form_supp}
\end{gather}
Therefore, when \(\mc\) acts on an arbitrary state two times, it should return the same state up to an overall phase factor \(e^{i \Omega}\) :
\begin{align}
    \mc^2 &= \mathcal{T} \cdot \sum_{\nu, \vn} e^{i \xi_\nu} \ketbra{f(\nu), -\vn - \vec{p}_\nu}{\nu, \vn} \cdot \mathcal{T} \cdot \sum_{\sigma, \vn'} e^{i \xi_\sigma} \ketbra{f(\sigma), -\vn' - \vec{p}_\sigma}{\sigma, \vn'} \notag \\
    & = \sum_{\nu, \vn, \sigma, \vn'} e^{-i \xi_\nu + i \xi_\sigma} \ketbra{f(\nu), -\vn - \vec{p}_\nu}{\sigma, \vn'} \delta_{\nu, f(\sigma)} \delta_{\vn, -\vn' - \vec{p}_\sigma} \notag \\
    & = \sum_{\sigma, \vn'} e^{-i \xi_{f(\sigma)} + i \xi_\sigma} \ketbra{\sigma, \vn' + \vec{p}_\sigma - \vec{p}_{f(\sigma)}}{\sigma, \vn'} = e^{i \Omega} \mathds{1}
\end{align}
where we used the fact that \(f\) is its own inverse: \(f(f(\sigma)) = \sigma\).
\(\mathds{1}\) is the identity operator both in the sublattice space as well as in the unit cell position space.
Therefore for all \(\sigma\)
\begin{gather}
    \xi_\sigma - \xi_{f(\sigma)} = \Omega \mod 2\pi, \notag \\
    \vec{p}_\sigma = \vec{p}_{f(\sigma)}.
    \label{eq:omega_condition}
\end{gather}
The condition \(\mc^2 \equiv \mathds{1}\) is equivalent to the particle-hole symmetry requirement (see Ref.~\cite{chiu2016classification} for details).
The Hamiltonian can be classified topologically according to the value of \(\Omega\), the presence of time-reversal and chiral symmetries.
If \(f(\sigma) = \sigma\) for at least one sublattice \(\sigma\) the condition Eq.~\eqref{eq:omega_condition} implies \(\Omega = 0\) and therefore \(\mathcal{A}^2 = \mathds{1}\).
This happens for example, for an odd number of sublattices.

\subsection{Conditions on the hopping matrices under anti-\(\pts\) symmetry of the Hamiltonian}
\label{app:anti_pt_cond_hoppings}

The Hamiltonian is given by
\begin{align}
    \mh = -\sum_{\vec{l}, \vn} \sum_{\nu, \sigma = 1}^{\mu} t_{\nu, \sigma}(\vec{l})\ketbra{\nu, \vn}{\sigma, \vn + \vec{l}} \quad \text{such that}~t^*_{\nu, \sigma}(\vec{l}) = t_{\sigma, \nu}(-\vec{l}).
\end{align}
Under the action of \(\mc\)~\eqref{eq:symmetry_operator_form_supp} we demand
\begin{align}
    & \mc \cdot \mh \cdot \mc^{-1} = - \mh \notag \\
    & \Rightarrow \sum_{\vec{l}, \vn} \sum_{\nu, \sigma = 1}^{\mu} e^{-i \xi_\nu + i \xi_\sigma} t^*_{\nu, \sigma}(\vec{l}) \ketbra{f(\nu), -\vn - \vec{p}_\nu}{f(\sigma), -\vn - \vec{l} - \vec{p}_\sigma}
    = -\sum_{\vec{l}, \vn} \sum_{\nu, \sigma = 1}^{\mu} t_{f(\nu), f(\sigma)}(\vec{l}) \ketbra{f(\nu), \vn}{f(\sigma), \vn + \vec{l}} \notag \\
    & \Rightarrow \sum_{\vec{l}, \vn} \sum_{\nu, \sigma = 1}^{\mu} e^{-i \xi_\nu + i \xi_\sigma} t^*_{\nu, \sigma}(\vec{l}) \ketbra{f(\nu), \vn}{f(\sigma), \vn - \vec{l} + \vec{p}_\nu - \vec{p}_\sigma}
    = - \sum_{\vec{l}, \vn} \sum_{\nu, \sigma = 1}^{\mu} t_{f(\nu), f(\sigma)}(\vec{l}) \ketbra{f(\nu), \vn}{f(\sigma), \vn + \vec{l}} \notag \\
    & \Rightarrow e^{-i \xi_\nu + i \xi_\sigma} t^*_{\nu, \sigma}(\vec{l}) = - t_{f(\nu), f(\sigma)}(- \vec{l} + \vec{p}_\nu - \vec{p}_\sigma)~.
    \label{eq:cond_hopp_antipt}
\end{align}
Note that the derivation of the condition~\eqref{eq:cond_hopp_antipt} did not impose any constraint on \(\vec{p}_\nu\): its components \(p_{\nu,i}\) can take any integer values.
Enforcing the anti-\(\pts\) symmetry on a lattice, restricts the possible values of the components of \(\vec{p}_\nu\) as was described in Section~\ref{sec:lattice_vec_antipt}.
However the above result suggests possible generalizations for Hamiltonians on generic graphs, provided a suitable generalization for the reflection operator is given.

\subsection{Anti-\(\pts\) symmetric generalized kagome lattice}
\label{app:kagome_cond}

We choose the following unit cell basis vectors for the 3D kagome lattice:
\begin{align}
    \vec{a}_1 = \hat{e}_1, \vec{a}_2 = \frac{1}{2} \hat{e}_1 + \frac{\sqrt{3}}{2} \hat{e}_2,~\vec{a}_3 = \hat{e}_3\;.
\end{align}
where \(\hat{e}_1, \hat{e}_2, \hat{e}_3\) are the Cartesian orthonormal coordinate axes.
Therefore we have \(\vec{n} = n_1 \vec{a}_1 + n_2 \vec{a}_2 + n_3 \vec{a}_3\), \(\vec{l} = l_1 \vec{a}_1 + l_2 \vec{a}_2 + l_3 \vec{a}_3\),
\(\vec{p}_1 = \vec{a}_2\), \(\vec{p}_2 = \vec{0}\), \(\vec{p}_3 = \vec{a}_1\), \(f(\nu) = \nu\).

The anti-\(\pts\) symmetry constrains the hopping parameters as implied by Eq.~\eqref{eq:cond_hopp_antipt}
\begin{align}\label{eq:apt_kagome_gen}
    & |t_{1,2}(0,0,0)| = |t_{1,2}(0,1,0)|, |t_{1,3}(0,0,0)| = |t_{3,1}(1,-1,0)|, \notag \\
    & |t_{2,3}(0,0,0)| = |t_{2,3}(-1,0,0)|, |t_{1,2}(0,0,1)| = |t_{1,2}(0,1,-1)|. \notag \\
    & \arg(t_{1,2}(0,1,0)) = \pi - \arg(t_{1,2}(0,0,0)) - \xi_1 + \xi_2~\text{mod}~2\pi, \notag \\
    & \arg(t_{3,1}(1,-1,0)) = \pi + \arg(t_{1,3}(0,0,0)) + \xi_1 - \xi_3~\text{mod}~2\pi, \notag \\
    & \arg(t_{2,3}(-1,0,0)) = \pi - \arg(t_{2,3}(0,0,0)) - \xi_2 + \xi_3~\text{mod}~2\pi, \notag \\
    & \arg(t_{1,2}(0,1,-1)) = \pi - \arg(t_{1,2}(0,0,1)) - \xi_1 + \xi_2~\text{mod}~2\pi,
\end{align}
and all the other hoppings equal zero.
We consider a special class of examples which satisfy the conditions~\eqref{eq:apt_kagome_gen} in the main text:
\begin{align}
    & |t_{1,2}(0,0,0)| = |t_{1,2}(0,1,0)| = t = 0.15, \notag \\
    & |t_{1,3}(0,0,0)| = |t_{3,1}(1,-1,0)| = |t_{2,3}(0,0,0)| = |t_{2,3}(-1,0,0)| = 1, \notag \\
    & |t_{1,2}(0,0,1)| = |t_{1,2}(0,1,-1)| = s. \notag \\
    & \arg(t_{3,1}(1,-1, 0)) = \pi, \arg(t_{2,3}(0,0,0)) = -\arg(t_{2,3}(-1,0,0)) = \varphi = \frac{\pi}{5}, \notag \\
    & \arg(t_{1,2}(0,1, 0)) = \arg(t_{1,3}(0,0,0)) = \arg(t_{1,2}(0,0,0)) = \arg(t_{1,2}(0,0,1)) = \arg(t_{1,2}(0,1,-1)) = 0~. 
    \label{eq:hopping_para_kagome_choice}
\end{align}
Diagonalizing the Hamiltonian \(\mh(\vk)\) with these choices of parameters \((t, \varphi, s)\) we obtain three bands:
\begin{gather}
    E = 0, \pm \big[2 s^2 \cos (k_2-2 k_3) + 4 t s \cos(k_2-k_3)+ 4 t s \cos(k_3) + 2 t^2 \cos(k_2) \notag \\
    + 2 \cos(k_1+2 \varphi)- 2 \cos (k_1-k_2) + 2 s^2 + 2 t^2 + 4 \big]^{\frac{1}{2}}.
\end{gather}
Therefore this choice of hoppings supports a flatband at \(E = 0\).
For the two dimensional kagome case we set \(s = 0\) and for the three dimensional case we set \(s = 2\).
The corresponding flatband eigenstate forms a compact localized state occupying five (three) unit cells in 3D (2D):
\begin{align}
    \frac{1}{\sqrt{4 + 2 t^2 + 2 s^2}}\Big[ & \ket{\nu = 1} \otimes \big\{e^{- i \varphi} \ket{n_1 + 1, n_2, n_3} + e^{i \varphi} \ket{n_1, n_2, n_3} \big\} \notag \\
    + & \ket{\nu = 2} \otimes \big\{\ket{n_1, n_2 + 1, n_3} - \ket{n_1 + 1, n_2, n_3} \big\} \notag \\
    - & \ket{\nu = 3} \otimes \big\{t \ket{n_1, n_2 + 1, n_3} + t \ket{n_1, n_2, n_3} + s \ket{n_1, n_2, n_3 + 1} + s \ket{n_1, n_2 + 1, n_3 - 1}\big\}\Big]~.
\end{align}

\subsection{Parametrization of DC field coordinates}
\label{sec:z_w_parametrization}

In this section we define the coordinates along and perpendicular to the DC field and present a way to parametrize these new coordinates.
We express the uniform DC field in a \(d\)-dimensional lattice in terms of the unit-cell basis vectors \(\vec{a}_j\):
\begin{gather}
    \vec{\mE} = \sum_{j = 1}^d \mE_j \vec{a}_j.
\end{gather}
We define the coordinate along the DC field as
\begin{gather}\label{eq:z_defin}
    z = \frac{1}{\mf} \vec{\mE} \cdot \vec{n} = \frac{1}{\mf}\sum_{j = 1}^d \mE_j \vec{a}_j \cdot \sum_{i = 1}^d n_{i} \vec{a}_i = \sum_{i = 1}^d n_i \mE_{\mf i},
    \qquad \mE_{\mf i} = \frac{1}{\mf} \sum_{j = 1}^d \mE_j \left(\vec{a}_j \cdot \vec{a}_i\right),
\end{gather}
where \(\mf\) is a proportionality factor to be specified later.
Similar to the unit cell coordinate along the field \(z\) we define the sublattice coordinate along the field:
\begin{align}\label{eq:z_nu_defin}
    z_\nu = \sum_{j = 1}^d m_{\nu, j} \mE_{\mf j}\;.
\end{align}

If the field direction is commensurate~\cite{mallick2021wannier} we can find \((d-1)\) vectors \(\{\vec{\mE}^\perp(s): s = 2, 3, \ldots, d\}\) perpendicular to the field,
and all of them can be expressed through unit-cell lattice vectors \(\vec{\mE}^\perp(s) = \sum_i \mE^\perp_i(s) \vec{a}_i\) where \(\mE^\perp_i(s) \in \mathbb{Z}\).
The corresponding orthogonality condition reads
\begin{align}
    \vec{\mE} \cdot \vec{\mE}^\perp(s) = 0 \Rightarrow \sum_{i = 1}^d \mE^\perp_{i}(s)\mE_{\mf i} = 0 \qquad \forall s.
    \label{supeq:percond}
\end{align}
Equation~\eqref{supeq:percond} is a set of \(d-1\) degenerate linear equations for \(d\) real variables \(\mE_{\mf i}\) with integer coefficients \(\mE^\perp_{i}(s)\).
Therefore one can fix one variable, for example \(\mE_{\mf 1}\), and determine all the other variables
as rational numbers \(\rho_i\),
\begin{align}
    \mE_{\mf i} = \rho_i \mE_{\mf 1}~\text{for}~i = 2, 3, \ldots, d.
    \label{eq:field_integer_con}
\end{align}
We assumed that \(\mE_{\mf 1}\neq 0\).
If \(\mE_{\mf 1} = 0\) then we can pick any other index \(j \neq 1\) for which \(\mE_{\mf j} \neq 0\) ---its existence is guaranteed for nonzero DC field.

Thanks to the relation~\eqref{eq:field_integer_con}, we can fix \(\mf\) by requiring \(\mE_{\mf i}\) to be integers for all \(i = 1, 2, 3, \ldots, d\) and enforcing \(\gcd(\mE_{\mf 1}, \mE_{\mf 2}, \mE_{\mf 3}, \ldots, \mE_{\mf d}) = 1\).
We conclude that \(z\) takes only integer values, furthermore because of the generalized B\'ezout's identity it takes \emph{all} integer values upon varying the lattice unit cell indices \(\{n_i\}\).
Equation~\eqref{eq:z_defin} is rewritten as
\begin{align}
    n_1 = \frac{1}{\mE_{\mf 1}} \left(z - \sum_{i = 2}^d n_i \mE_{\mf i}\right).
    \label{eq:n_choice}
\end{align}

Using the relation~\eqref{eq:n_choice} the coordinates along \(\vec{\mE}^\perp(s)\) are expressed as a function of the \(z\) coordinate and other parameters \(n_{i>1}\):
\begin{align}
    \frac{1}{\mf} \vn \cdot \vec{\mE}^\perp(s) = \sum_{j = 1}^d n_j \mE^\perp_{\mf j}(s) = \frac{z \mE^\perp_{\mf 1}(s)}{\mE_{\mf 1}} + \sum_{j = 2}^d n_j
    \left(\mE^\perp_{\mf j}(s) - \mE_{\mf j} \mE^\perp_{\mf1}(s) /\mE_{\mf 1}\right),
\end{align}
where we defined \(\mE^\perp_{\mf i}(s) = \frac{1}{\mf}\sum_{j = 1}^d \mE^\perp_{j}(s) \vec{a}_j \cdot \vec{a}_i\).
We define a rescaled coordinate \(w(s)\) along \(\vec{\mE}^\perp(s)\) by multiplying the above equation with a factor \(\mE_{\mf 1}\):
\begin{align}
    w(s) = \frac{\mE_{\mf 1}}{\mf} \vn \cdot \vec{\mE}^\perp(s) = w_z(s) + \sum_{j = 2}^d n_j \Delta_j\; \text{where}~
    \Delta_j = \mE^\perp_{\mf j}(s)\mE_{\mf 1} - \mE_{\mf j} \mE^\perp_{\mf 1}(s), w_z(s) = z \mE^\perp_{\mf 1}(s).
    \label{eq:w_paramet}
\end{align}
For a fixed \(z\) all the allowed values of \(w(s)\) form a periodic lattice structure.
Note that unlike \(z\) the perpendicular coordinates \(w(s)\) are not integer in general, and cannot be made integer by applying suitable scaling factors~\cite{mallick2021wannier}.
Because of the \(\Delta_j\) factor the physical distance between the two neighboring \(w(s)\) coordinates for a fixed \(s\) and \(z\) is in general different from the nearest neighbor distance for the original unit cell coordinate \(\vn\).

We use independent coordinates \((z, \eta_2, \eta_3, \ldots, \eta_d)\) to parametrize the \(w(s)\) set.
Here the integer components \(\eta_j\) are the equivalents of \(n_j\) in the above expressions.
We also use these integer coordinates to label the Hilbert space basis \(\ket{\nu, \vn} = \ket{\nu, z, \vec{\eta}}\), \(\vec{\eta}\) is a \((d-1)\)-dimensional vector with integer components.

\subsection{The Hamiltonian in the presence of the DC field}
\label{app:Hamil_in_dc_field}

The tight-binding single-particle Hamiltonian in the presence of a uniform DC field is
\begin{align}
    \mhme = \vmE \cdot \hat{r} - \sum_{\vec{l}, \vn, \nu, \sigma} t_{\nu, \sigma}(\vec{l}) \ketbra{\nu, \vn}{\sigma, \vn + \vec{l}}.
\end{align}
Using Eqs.~\eqref{eq:z_defin} and~\eqref{eq:z_nu_defin} the potential energy term of the Hamiltonian can be simplified as follows
\begin{align}
    \vec{\mE} \cdot \hat{r} = \mf \sum_{z, \vec{\eta}}
    \sum_{\nu = 1}^\mu \sum_{j = 1}^d (n_j + m_{\nu, j}) \mE_{\mf j} \ketbra{\nu, z, \vec{\eta}}{\nu, z, \vec{\eta}}
    = \mf \sum_{z, \vec{\eta}} \sum_{\nu = 1}^\mu (z + z_\nu) \ketbra{\nu, z, \vec{\eta}}{\nu, z, \vec{\eta}}.
\end{align}
The action of the hopping shifts lattice vector components from \(n_i\) to \(n_i - l_i\), where we expanded the hopping vector over the unit cell basis vectors: \(\vec{l} = \sum_i l_i \vec{a}_i\).
Therefore the total Hamiltonian expressed in the coordinates \(z,\vec{\eta}\) reads as
\begin{align}
    \mhme = \sum_{\nu = 1}^\mu \sum_{z, \vec{\eta}} \left[\mf (z + z_\nu) \ketbra{\nu, z, \vec{\eta}}{\nu, z, \vec{\eta}}
    - \sum_{\vec{l}} \sum_{\sigma = 1}^\mu t_{\nu, \sigma}(\vec{l}) \ketbra{\nu, z - \sum_{j = 1}^d l_j \mE_{\mf j}, \vec{\eta} - \vec{\epsilon}(\vec{l})}{\sigma, z, \vec{\eta}} \right]\;. \label{eq_ws_hamil_rotated_coord}
\end{align}
Here \(\vec{\epsilon}(\vec{l}) = (\epsilon_2(\vec{l}), \epsilon_3(\vec{l}), \ldots, \epsilon_d(\vec{l}))\) is a \(d-1\) dimensional vector which for our parametrization---e.g., for the choices made in Eqs.~(\ref{eq:n_choice}) and (\ref{eq:w_paramet})---becomes \((l_2, l_3, \ldots, l_d)\).
A different parametrization compared to the case described in Eqs.~\eqref{eq:n_choice} and~\eqref{eq:w_paramet} might change the form of the \(d-1\) dimensional vectors \(\vec{\epsilon}\).
For a general parametrization \(\vec{\epsilon}\) is a linear function of all the \(d\) components \((l_1, l_2, \ldots, l_d)\).
The results presented below are valid irrespective of the choice of the parametrization.

\subsection{Diagonalization of the Hamiltonian in the presence of a DC field}
\label{sec:effective_WS_hamil}

The Hamiltonian \(\mhme\) is translationally invariant in \(\vec{\eta}\), therefore we can apply the Bloch's theorem and partially diagonalize it using the Fourier transform over \(\vec{\eta}\)
\begin{align}
    \ket{\phi(\nu, z, \vec{\kappa})} = \frac{1}{{(2\pi)}^{\frac{d-1}{2}}}\ket{\nu, z} \otimes \sum_{\vec{\eta}} e^{i \vec{\kappa} \cdot \vec{\eta}} \ket{\vec{\eta}},
    \qquad \vec{\kappa} \cdot \vec{\eta} = \sum_{j = 2}^d \kappa_j \eta_j\;.
\end{align}
In the Bloch basis the Hamiltonian is block diagonal, and each block corresponds to a fixed \(d-1\) dimensional momentum \(\vec{\kappa}\) and acts on the \(z\) and sublattice spaces:
\begin{align}
    & \mhme = \int_{\vec{\kappa}} d\vec{\kappa}~ \mhme(\vec{\kappa}), \notag \\
    & \mhme(\vec{\kappa}) = \sum_{z,\nu} \left[\mf (z + z_\nu) \ketbra{\phi(\nu, z, \vec{\kappa})}{\phi(\nu, z, \vec{\kappa})}
    - \sum_{\vec{l}, \sigma} t_{\nu, \sigma}(\vec{l}) e^{i\vec{\kappa}\cdot \vec{\epsilon}(\vec{l})}
    \ketbra{\phi(\nu,z - \sum_{j = 1}^d l_j \mE_{\mf j}, \vec{\kappa})}{\phi(\sigma, z, \vec{\kappa})}\right].
    \label{eq:hamil_in_kappa}
\end{align}

The eigenvector of \(\mhme\) with eigenvalue \(E\) is a function of \(\vec{\kappa}\) only:
\begin{gather}
    \mhme(\vec{\kappa}) \ket{\psi_E(\vec{\kappa})} = E \ket{\psi_E(\vec{\kappa})}\;. \label{eq:eigen_eq_supp}
\end{gather}
It is expressed as a linear superposition of wavevectors \(\ket{\phi(\nu, z, \vec{\kappa})}\) over different \(z\) and \(\nu\):
\begin{gather*}
    \ket{\psi_E(\vec{\kappa})} = \sum_{\nu, z} \psi_E(\nu, z, \vec{\kappa}) \ket{\phi(\nu, z, \vec{\kappa})},
\end{gather*}
so that the eigenvalue Eq.~\eqref{eq:eigen_eq_supp} turns 
\begin{gather}
    \left[\mf (z + z_\nu) - E\right] \psi_E(\nu, z, \vec{\kappa}) = \sum_{\vec{l}, \sigma} t_{\nu, \sigma}(\vec{l}) e^{i\vec{\kappa}\cdot \vec{\epsilon}(\vec{l})}
    \psi_E\left(\sigma, z + \sum_{i = 1}^d l_i \mE_{\mf i}, \vec{\kappa}\right).
    \label{eq:eige_prob}
\end{gather}
This is a system of linear equations that couple different values of \(z\).
For a given \(z\) this system is a set of \(\mu\) linear coupled equations.
Because of the linearity of the system we can define the following generating function (a Fourier transformation over \(z\)):
\begin{align}
    g_E(\nu, q, \vec{\kappa}) = \frac{1}{\sqrt{2\pi}} e^{-\frac{i E q}{\mf}} \sum_{z \in \mathbb{Z}} e^{i q (z + z_\nu)} \psi_E(\nu, z, k),
\end{align}
which is possible since \(z\) takes only integer values for a commensurate DC field:
This turns the eigensystem~\eqref{eq:eige_prob} into a set of \(\mu\) coupled differential equations
\begin{align}
    \label{eq:gen_evl}
    i\frac{\partial}{\partial q} \vec{g}_E(q, \vec{\kappa}) & = \mhme(q, \vec{\kappa}) \cdot \vec{g}_E(q, \vec{\kappa}), \notag\\
    \mhme(q, \vec{\kappa}) & = -\frac{1}{\mf} \sum_{\vec{l}, \nu, \sigma} e^{i q(z_\nu - z_\sigma)} t_{\nu \sigma}(\vec{l})
    e^{i \vec{\kappa} \cdot \vec{\epsilon}(\vec{l})-i q \sum_{i = 1}^d l_i \mE_{\mf i}} \ketbra{\nu}{\sigma}, \notag\\
    \vec{g}_E(q, \vec{\kappa}) & = [g_E(1, q, \vec{\kappa})~g_E(2, q, \vec{\kappa})~\ldots g_E(\mu, q, \vec{\kappa})]^T \;.
\end{align}
Equation~\eqref{eq:gen_evl} is a Schr\"odinger equation describing unitary evolution under the effective Hermitian Hamiltonian \(\mhme(q, \vec{\kappa})\), but with time \(t\) replaced by the variable \(q\).
Therefore the \(\vec{g}_E(q, \vec{\kappa})\) evolve unitarily in \(q\)-space:
\begin{align}
    \vec{g}_E(q, \vec{\kappa}) = U(q, \vec{\kappa}) \cdot \vec{g}_E(0, \vec{\kappa})
    \label{eq:gen_evl_unitary}\;.
\end{align}
Here \(U(q, \vec{\kappa})\) is the \(q\)-ordered exponential of the integrated matrix \(\mhme(q, \vec{\kappa})\):
\begin{align}
    & U(q, \vec{\kappa}) = \mathds{1} + (- i) \int_{q' = 0}^q dq' \mhme(q', \vec{\kappa})
    + {(-i)}^2 \int_{q' = 0}^q \int_{q'' = 0}^{q'} dq' dq'' \mhme(q', \vec{\kappa}) \mhme(q'', \vec{\kappa}) + \ldots
    \label{con:qspace_unit_evol}
\end{align}
The generating function satisfies
\begin{align}
    & g_E(\nu, q + 2\pi, \vec{\kappa}) = e^{-\frac{2 \pi i E}{\mf}} e^{2 \pi i z_\nu} g_E(\nu, q, \vec{\kappa})\;.
    \label{eq:gen_period}
\end{align}
From the above and Eq.~\eqref{eq:gen_evl_unitary} we get
\begin{align}
    \vec{g}_E(2\pi, \vec{\kappa}) = U(2\pi, \vec{\kappa}) \cdot \vec{g}_E(0, \vec{\kappa}) & = e^{-\frac{2 \pi i E}{\mf}} \Lambda(2\pi) \cdot \vec{g}_E(0, \vec{\kappa}) \notag \\
   \Rightarrow \big[\Lambda^\dagger(2\pi) \cdot U(2\pi, \vec{\kappa})\big] \cdot \vec{g}_E(0, \vec{\kappa}) & = e^{-\frac{2 \pi i E}{\mf}} \vec{g}_E(0, \vec{\kappa})\;.
    \label{eq:gen_eigen}
\end{align}
Here we defined the diagonal matrix associated with the potential energy at sublattice sites:
\begin{align}
    \Lambda(q) = \sum_{\nu = 1}^{\mu} e^{i q z_\nu}\ketbra{\nu}{\nu}\;.
    \label{eq:onsite_poten_sublattice}
\end{align}
Equation~\eqref{eq:gen_eigen} is an eigenvalue equation for the operator \(\Lambda^\dagger(2\pi) \cdot U(2\pi, \vec{\kappa})\) which has \(\mu\) possible eigenvalues \(e^{-\frac{2 \pi i E}{\mf}}\).
These are functions of \(\vec{\kappa}\) only (because \(E\) depends on \(\vec{\kappa}\) only).
At the same time, for each \(\vec{\kappa}\) the total number of eigenvalues of \(\mhme(\vec{\kappa})\) must be \(\mu~\times\) the total number of values that is taken by the coordinate \(z\).
All the eigenvalues of the Hamiltonian \(\mhme(\vec{\kappa})\) are generated using the periodicity of \(e^{-\frac{2 \pi i E}{\mf}}\):
\begin{align}\label{eig_val}
    E_{\gamma,a}(\vec{\kappa}) = \mf a + \frac{i \mf}{2\pi} \ln\left[\lambda_{\gamma}(k)\right] \text{where}~\gamma = 1, 2, \ldots,\mu;~a \in \mathbb{Z},
\end{align}
where \(\lambda\) are the eigenvalues of the \(\mu \times \mu\) unitary matrix \(\Lambda^\dagger(2\pi) \cdot U(2\pi, \vec{\kappa})\).
The two subscripts \(\gamma\) and \(a\) of the eigenvalues \(E\) identify the Wannier-Stark bands.
For any \(a\), \(E_{\gamma,a}(\vec{\kappa})\) is an irreducible band structure and carries the complete information about the band structure.
For any \(b \neq a\), \(E_{\gamma, b}(\vec{\kappa})\) can be generated from \(E_{\gamma,a}(\vec{\kappa})\) by a constant shift  \(\mf (b - a)\) in energy.

\subsection{Robustness of anti-\(\pts\) symmetric flatbands in the presence of a DC field}
\label{app:anti_pt_under_dc_field_proof}

From the condition~\eqref{cond:antipt_lattice_pos}
we obtain
\begin{align}
    z_{\nu} + z_{f(\nu)} = \sum_{i = 1}^d p_{\nu, i} \mE_{\mf i}\;.
    \label{con:antipt_z}
\end{align}
Therefore since \(p_{\nu, i}\) is integer by the definition of a lattice vector, and since \(\mE_{\mf i}\) is integer because of the choice of the commensurate DC field direction,
we conclude that \(z_{\nu} + z_{f(\nu)}\) can only take integer values.
The unitary matrix \(\Lambda(q)\) is diagonal, therefore \(\Lambda^* (q) = \Lambda^\dagger(q)\).
The \(d-1\) dimensional vector \(\vec{\epsilon}\) defined in Eq.~\eqref{eq_ws_hamil_rotated_coord} is a linear function of \(\vec{l}\) and therefore
\begin{align}
    \vec{\epsilon}(- \vec{l} + \vec{p}_\nu - \vec{p}_\sigma) = -\vec{\epsilon}(\vec{l}) + \vec{\epsilon}(\vec{p}_\nu) - \vec{\epsilon}(\vec{p}_\sigma)\;.
    \label{con:shift_wspace}
\end{align}
From Eqs.~\eqref{eq:gen_evl} and~\eqref{eq:onsite_poten_sublattice} we obtain
\begin{align*}
    \mhme(q, \vec{\kappa}) & = \Lambda(q) \cdot \sum_{\vec{l}} T(\vec{l}) \cdot \Lambda^* (q)\;, \\
    T(\vec{l}) & = -\frac{1}{\mf} \sum_{\nu, \sigma} t_{\nu,\sigma}(\vec{l}) e^{i \vec{\kappa} \cdot \vec{\epsilon}(\vec{l})-i q \sum_{i = 1}^d l_i \mE_{\mf i}} \ketbra{\nu}{\sigma}\;.
\end{align*}
Using the anti-\(\pts\) symmetry conditions for the hopping parameters~\eqref{eq:cond_hopp_antipt} and for the \(z_\nu\) coordinate~\eqref{con:antipt_z}, together with the condition~\eqref{con:shift_wspace} we write
\begin{align}
    \sum_{\vec{l}} T^* (\vec{l}) &= \frac{1}{\mf} \sum_{\vec{l}, \nu, \sigma} e^{i \xi_\nu - i \xi_\sigma}
    t_{f(\nu),f(\sigma)}(-\vec{l} + \vec{p}_\nu + \vec{p}_\sigma) e^{-i \vec{\kappa} \cdot \vec{\epsilon}(\vec{l})+ i q \sum_{i = 1}^d l_i \mE_{\mf i}} \ketbra{\nu}{\sigma} \notag \\
    & = \frac{1}{\mf} \sum_{\vec{l}, \nu, \sigma} e^{i \xi_\nu - i \xi_\sigma}
    t_{f(\nu),f(\sigma)}(\vec{l}) e^{i \vec{\kappa} \cdot \vec{\epsilon}(\vec{l})- i q \sum_{i = 1}^d l_i \mE_{\mf i}}
    e^{-i \vec{\kappa} \cdot \vec{\epsilon}(\vec{p}_\nu) + i q \sum_{i = 1}^d p_{\nu, i} \mE_{\mf i}}
    e^{i \vec{\kappa} \cdot \vec{\epsilon}(\vec{p}_\sigma) - i q \sum_{i = 1}^d p_{\sigma, i} \mE_{\mf i}} \ketbra{\nu}{\sigma} \notag \\
    & = \frac{1}{\mf} \sum_{\vec{l}, \nu, \sigma} e^{i \xi_\nu - i \xi_\sigma}
    t_{f(\nu),f(\sigma)}(\vec{l}) e^{i \vec{\kappa} \cdot \vec{\epsilon}(\vec{l})- i q \sum_{i = 1}^d l_i \mE_{\mf i}}
    e^{-i \vec{\kappa} \cdot \vec{\epsilon}(\vec{p}_\nu) + i q (z_\nu + z_{f(\nu)})}
    e^{i \vec{\kappa} \cdot \vec{\epsilon}(\vec{p}_\sigma) - i q (z_\sigma + z_{f(\sigma)})} \ketbra{\nu}{\sigma} \notag \\
    & = \Lambda(q) \cdot \mathcal{M}^\dagger(\vec{\kappa}) \cdot \Lambda(q) \cdot
    \left[\frac{1}{\mf} \sum_{\vec{l}, \nu, \sigma} t_{f(\nu),f(\sigma)}(\vec{l}) e^{i \vec{\kappa} \cdot \vec{\epsilon}(\vec{l})- i q \sum_{i = 1}^d l_i \mE_{\mf i}}
    \ketbra{f(\nu)}{f(\sigma)} \right]\cdot \Lambda^* (q) \cdot \mathcal{M}(\vec{\kappa}) \cdot \Lambda^* (q) \notag \\
    & = -\Lambda(q) \cdot \mathcal{M}^\dagger(\vec{\kappa}) \cdot \Lambda(q) \cdot \sum_{\vec{l}} T(\vec{l}) \cdot \Lambda^* (q) \cdot \mathcal{M}(\vec{\kappa}) \cdot \Lambda^* (q), \\
    \mathcal{M}(\vec{\kappa}) & = \sum_{\nu} e^{-i \xi_\nu}e^{i \vec{\kappa} \cdot \vec{\epsilon} (\vec{p}_\nu)} \ketbra{f(\nu)}{\nu}\;. \notag
\end{align}
Complex conjugation of the Hamiltonian \(\mhme(q, \vkp)\) results in
\begin{align}
    & \mhme^* (q, \vec{\kappa}) = \sum_{\vec{l}} \Lambda^* (q) \cdot T^* (\vec{l}) \cdot \Lambda(q) = -\mathcal{M}^\dagger(\vec{\kappa}) \cdot \Lambda(q) \cdot \sum_{\vec{l}} T(\vec{l}) \cdot \Lambda^* (q) \cdot \mathcal{M}(\vec{\kappa})
    = - \mathcal{M}^\dagger(\vec{\kappa}) \cdot \mhme(q, \vec{\kappa}) \cdot \mathcal{M}(\vec{\kappa})\;.
    \label{con:reduced_anti_uni}
\end{align}
Since \(\mathcal{M}\) is a unitary matrix, the Hamiltonian \(\mhme(q, \vec{\kappa})\) is anti-symmetric under an antiunitary operation.
From the definition of \(\Lambda\) and Eq.~\eqref{con:antipt_z} we obtain:
\begin{align}
    \Lambda^\dagger(2\pi) = \mathcal{M}(\vec{\kappa}) \cdot \Lambda(2\pi) \cdot \mathcal{M}^{\dagger}(\vec{\kappa})\;.
    \label{con:lambda_m_2pi}
\end{align}
Using the condition~\eqref{con:reduced_anti_uni} in the series expansion of the evolution operator~\eqref{con:qspace_unit_evol} we obtain
\begin{align}
    U^* (q, \vec{\kappa}) = \mathcal{M}^{\dagger}(\vec{\kappa}) \cdot U(q, \vec{\kappa}) \cdot \mathcal{M}(\vec{\kappa})\;.
    \label{con:on_unit_mk}
\end{align}
Now the complex conjugation of the eigenvalue equation for the generating function~\eqref{eq:gen_eigen} becomes
\begin{align}
    \big[\Lambda(2\pi) \cdot U^* (2\pi, \vec{\kappa})\big] \cdot \vec{g}^*_E(0, \vec{\kappa}) & = e^{\frac{2 \pi i E}{\mf}} \vec{g}^*_E(0, \vec{\kappa}) \notag \\
    \Rightarrow \big[\Lambda(2\pi) \cdot \mathcal{M}^\dagger(\vec{\kappa}) \cdot U(2\pi, \vec{\kappa}) \cdot \mathcal{M}(\vec{\kappa})\big] \cdot \vec{g}^*_E(0, \vec{\kappa})
    & = e^{\frac{2 \pi i E}{\mf}} \vec{g}^*_E(0, \vec{\kappa}) \notag \\
    \Rightarrow \Lambda^\dagger(2\pi) \cdot U(2\pi, \vec{\kappa}) \cdot \big[\mathcal{M}(\vec{\kappa})\cdot \vec{g}^*_E(0, \vec{\kappa})\big]
    & = e^{\frac{2 \pi i E}{\mf}} \big[\mathcal{M}(\vec{\kappa})\cdot \vec{g}^*_E(0, \vec{\kappa})\big]\;.
    \label{eq:gen2_eigen}
\end{align}
Here we used the conditions~\eqref{con:lambda_m_2pi} and~\eqref{con:on_unit_mk}.

Comparing Eqs.~\eqref{eq:gen_eigen} and~\eqref{eq:gen2_eigen} we conclude that within each irreducible Wannier-Stark band structure for each eigenvalue \(e^{-\frac{2 \pi i E}{\mf}}\) of the unitary operator \(\Lambda^\dagger(2\pi) \cdot U(2\pi, \vec{\kappa})\) there exists another eigenvalue \(e^{\frac{2 \pi i E}{\mf}}\).
If the number of sublattices per unit cell \(\mu\) is odd, then there exists at least one eigenvalue such that \(e^{-\frac{2 \pi i E}{\mf}} = e^{\frac{2 \pi i E}{\mf}}\), implying \( E(\vkp) = 0 \mod \frac{\mf}{2}\),
which implies in turn a constant \(E\) for all \(\vec{\kappa}\), e.g., a flatband in the presence of a DC field.


\subsection{Reproducing the anti-\(\pts\) Wannier-Stark band structure with a Floquet Hamiltonian}
\label{supp:Floquet_mapp}

Here we demonstrate how a band structure of an anti-\(\pts\) symmetric Hamiltonian on a \(d\)-dimensional tight-binding network can be reproduced by a Floquet Hamiltonian on a \((d-1)\)-dimensional tight-binding network with time-periodic Peierls phases.
This provides a possibility for an experimental implementations of our theory in the state-of-art setups.
For example, see the experiments in Refs.~\cite{{bordia2017periodically, eckardt2017colloquium}}, which implement Floquet (periodic in time) Hamiltonians using ultracold atoms.

The starting point is a driven Hamiltonian on a \((d-1)\)-dimensional non-Bravais lattice with time-periodic hopping parameters:
\begin{align}
    \mh(\tau) = -\sum_{\nu, \sigma = 1}^\mu \sum_{\vec{l}} t_{\nu\sigma}(\vec{l}) e^{-i \tau \mf \sum_{j = 1}^d l_j \mE_{\mf j}} \ketbra{\nu}{\sigma} \otimes \sum_{\vec{\eta}}\ketbra{\vec{\eta} - \vec{\epsilon}}{\vec{\eta}} 
    + \mf \sum_{\nu = 1}^\mu z_\nu \ketbra{\nu}{\nu} \otimes \sum_{\vec{\eta}}\ketbra{\vec{\eta}}{\vec{\eta}}\;. 
    \label{eq:fl_ham}
\end{align}
Here \(\tau\) is time; \((d-1)\)-dimensional vector \(\vec{\epsilon}\) is a linear function of hopping vector \(\vec{l}\) as defined in Eq.~\eqref{eq_ws_hamil_rotated_coord};
\(\mf z_\nu\) is a potential energy at sublattice \(\nu\), independent of the unit cell \(\vec{\eta}\).
The associated Hilbert space \(\equiv \mathbb{C}^\mu \otimes \mathbb{C}^{d-1}\) is spanned by \(\{\ket{\nu, \vec{\eta}}:~\nu = 1,2,\ldots, \mu; ~\text{components of}~\vec{\eta} \in \mathbb{Z}^{d-1}\}\).
The Peierls phase parameter \(\sum_{j = 1}^d l_j \mE_{\mf j}\) has the same form as in Eq.~\eqref{eq_ws_hamil_rotated_coord} and consequently it is always an integer.
Therefore the time period of the Hamiltonian is \(\frac{2\pi}{\mf}\),
\begin{align}
    \mh(\tau + 2\pi/\mf) = \mh(\tau),
\end{align}
and \(\mf\) is a circular frequency.
Using Floquet's Theorem we can write the eigenfunction as a product of a time-periodic wavefunction having the same period of the Hamiltonian and an additional phase factor \(e^{-i E \tau}\):
\begin{align}
    \ket{\Phi_E(\tau)} = e^{-i E \tau} \sum_{z, \nu, \vec{\eta}} e^{iz\mathcal{F} \tau} \Phi_E(\nu, z, \vec{\eta}) \ket{\nu,\vec{\eta}}\;.
\end{align}
Here the time-periodic part of the wavefunction is expanded in a Fourier series.
In this case the number \(z \in \mathbb{Z}\) corresponds to frequencies, not spatial coordinates.
However, thanks to the above representation of an eigenstate, the time-dependent Schr\"odinger equation becomes
\begin{align}
    & i \frac{\partial}{\partial \tau}\ket{\Phi_E(\tau)} = \mh(\tau) \ket{\Phi_E(\tau)} \notag \\
    \Rightarrow &\quad \sum_{\nu, z, \vec{\eta}} (E - z \mf)e^{i z \mf \tau} \Phi_E(\nu, z, \vec{\eta}) \ket{\nu,\vec{\eta}}
    = -\sum_{\nu, \sigma} \sum_{\vec{l}, \vec{\eta}, z} t_{\nu\sigma}(\vec{l}) e^{-i \tau \mathcal{F} \left(\sum_{j = 1}^d l_j \mE_{\mf j} - z\right)} \Phi_E(\sigma, z, \vec{\eta})\ket{\nu, \vec{\eta} - \vec{\epsilon}}
    \notag \\
    &\quad \quad \quad \quad \quad \quad \quad \quad \quad \quad \quad \quad \quad \quad \quad \quad \quad \quad \quad \quad \quad \quad \quad \quad \quad \quad \quad 
    + \mf \sum_{z, \nu} \sum_{\vec{\eta}} z_\nu e^{i z \mf \tau} \Phi_E(\nu, z, \vec{\eta}) \ket{\nu,\vec{\eta}}\;.
    \label{eq:schr_static}
\end{align}
This is transformed into a static Schr\"odinger equation by looking at equations for individual Fourier components in the above equation, e.g., for fixed \(z\), \(\nu\) and \(\vec{\eta}\)
\begin{align}
    (E - z \mf)\Phi_E(\nu, z, \vec{\eta})
    = -\sum_{\sigma} \sum_{\vec{l}} t_{\nu\sigma}(\vec{l})~\Phi_E\bigg(\sigma, z + \sum_{j = 1}^d l_j \mE_{\mf j}, \vec{\eta} + \vec{\epsilon}\bigg)
    + \mf z_\nu \Phi_E(\nu, z, \vec{\eta})\;.
 \end{align}
Introducing a new extended Hilbert space spanned additionally by frequency multiples: \(\{\ket{z}\}\), we can rewrite the above eigen-equation as
\begin{align}
    \left[\sum_{\nu, z, \vec{\eta}} \mf (z + z_\nu)\ketbra{\nu, z, \vec{\eta}}{\nu, z, \vec{\eta}} -
    \sum_{\nu, \sigma}\sum_{\vec{l}} t_{\nu\sigma}(\vec{l}) \ketbra{\nu}{\sigma} \otimes \sum_{z, \vec{\eta}} \ketbra{z - \sum_{j = 1}^d l_j \mE_{\mf j}, \vec{\eta} - \vec{\epsilon}}{z, \vec{\eta}}
    \right] \cdot \ket{\Phi_E} = E \ket{\Phi_E}, \notag
\end{align}
with the redefinition of the eigenstate in the extended Hilbert space
\begin{gather*}
    \ket{\Phi_E} = \sum_{\nu, z, \vec{\eta}} \Phi_E(\nu, z, \vec{\eta}) \ket{\nu, z, \vec{\eta}}.
\end{gather*}

The above equation is identical to a time-independent Schr\"odinger equation for a static Hamiltonian on a \(d\)-dimensional lattice with DC field having strength \(\mf\) (see Eq.~\eqref{eq_ws_hamil_rotated_coord} for further details)
\begin{align}
    \mhme = \sum_{\nu = 1}^\mu \sum_{z, \vec{\eta}} \left[\mf (z + z_\nu) \ketbra{\nu, z, \vec{\eta}}{\nu, z, \vec{\eta}}
    - \sum_{\vec{l}} \sum_{\sigma = 1}^\mu t_{\nu, \sigma}(\vec{l}) \ketbra{\nu, z - \sum_{j = 1}^d l_j \mE_{\mf j}, \vec{\eta} - \vec{\epsilon}}{\sigma, z, \vec{\eta}} \right]\;.
    \label{eq_ws_hamil_rotated_coord_2}
\end{align}
Therefore both Hamiltonians of Eqs.~\eqref{eq:fl_ham} and~\eqref{eq_ws_hamil_rotated_coord_2} have the same spectrum.
Note that we need the same number of sublattices \(\mu\) per unit cell for both the \(d\)-dimensional Wannier-Stark problem and the \((d-1)\)-dimensional Floquet problem.
It is important to point out, that the repeating set of Wannier-Stark bands folds into a single set of Floquet bands, which are periodic in energy.

\subsubsection{Example: 2D and 3D kagome lattices}

In the main text we introduced the tight-binding anti-\(\pts\) Hamiltonians for the kagome lattice and its 3D version, which have Wannier-Stark flatbands.
From the above derivation it follows that the same spectrum can be obtained for Floquet Hamiltonians on 1D and 2D diamond lattices, respectively---with an appropriate choice of hopping parameters.
We use the same parametrization of field coordinates as in the previous sections: assuming \(\mE_{\mf 1} \neq 0\) we find \(\epsilon_2 = l_2\), \(\epsilon_3 = l_3\).

\begin{figure}
    \includegraphics[width = 0.4\textwidth]{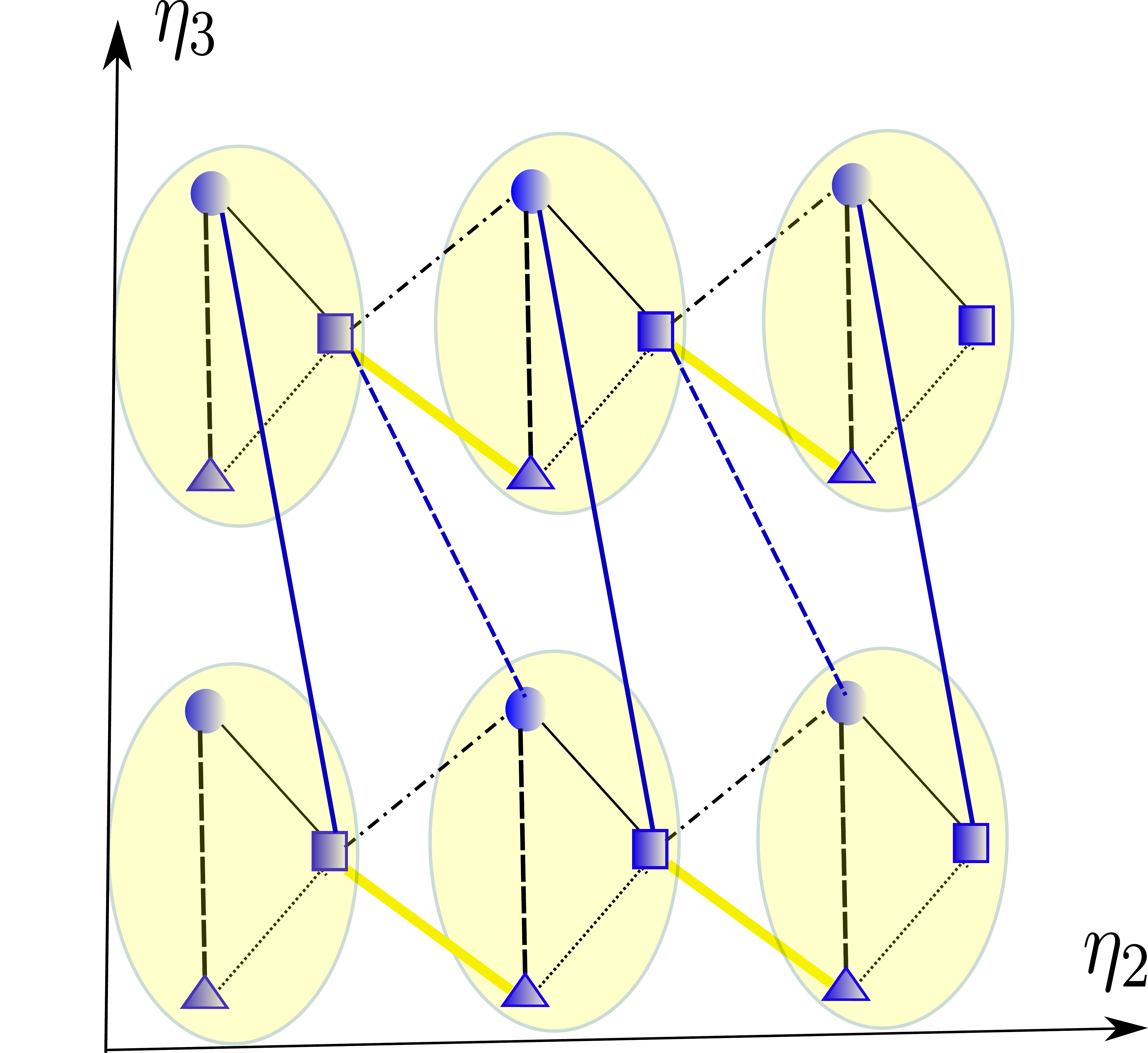}
    \caption{
Part of the 2D diamond lattice containing six unit cells. Unit cells are shown in yellow shaded ellipses.
        Square shaped sites denote the first sublattice, sphere shaped sites denote the second sublattice,
        and triangle shaped sites denote the third sublattice. They are represented, respectively, by the vectors \(\ket{1}\), \(\ket{2}\) and \(\ket{3}\) in the Floquet Hamiltonian \eqref{eq:fl_ham_diamond_2D}. The potential energies at first, second and third sublattice sites are \(\frac{\mf}{2}\mE_{\mf 2}\),  \(0\) and  \(\frac{\mf}{2}\mE_{\mf 1}\) respectively, for any unit cell.
        Hopping parameters within a unit cell: black solid thin line \(\equiv -t\), black dashed line \(\equiv -e^{i \varphi} - e^{-i \varphi + i \tau \mf \mE_{\mf 1}}\),
        black dotted line \(\equiv -1\); hopping between unit cells along \(\eta_2\) axis: black dash-dotted line \(\equiv -t e^{-i \tau \mf \mE_{\mf 2}}\),
        yellow thick solid line \(\equiv  e^{i \tau \mf(\mE_{\mf 1} - \mE_{\mf 2})}\); hopping between unit cells along \(\eta_3\) axis: blue solid line \(\equiv -s e^{-i \tau \mf \mE_{\mf 3}}\),
        blue dashed line \(\equiv -s e^{-i \tau \mf (\mE_{\mf 2} - \mE_{\mf 3})} \). See the Hamiltonian in Eq.~\eqref{eq:fl_ham_diamond_2D} for more details.}
    \label{fig:2d_diamond}
\end{figure}

It is convenient to start with the 2D diamond lattice case.
The unit cell coordinates are indexed by \(\vec{\eta}\) = \((\eta_2, \eta_3)\).
From the Hamiltonian~\eqref{eq:hopping_para_kagome_choice} we can write the Floquet Hamiltonian [see Eq.~\eqref{eq:fl_ham}]
{\small
\begin{align}
    \mh_{2D}(\tau)
    = -\Big[t \ketbra{1}{2} + t \ketbra{2}{1} + \big\{e^{i \varphi} + e^{-i \varphi + i \tau \mf \mE_{\mf 1}}\big\}\ketbra{2}{3} + \big\{e^{-i \varphi} + e^{i \varphi - i \tau \mf \mE_{\mf 1}}\big\} \ketbra{3}{2} + \ketbra{1}{3} + \ketbra{3}{1}\Big] \otimes \sum_{\eta_2, \eta_3}\ketbra{\eta_2, \eta_3}{\eta_2, \eta_3} \notag\\
    - \Big[t e^{-i \tau \mf \mE_{\mf 2}} \ketbra{1}{2} - e^{i \tau \mf(\mE_{\mf 1} - \mE_{\mf 2})}\ketbra{1}{3}\Big] \otimes \sum_{\eta_2, \eta_3}\ketbra{\eta_2 - 1, \eta_3}{\eta_2, \eta_3} \notag\\
    - \Big[t e^{i \tau \mf \mE_{\mf 2}} \ketbra{2}{1} - e^{-i \tau \mf(\mE_{\mf 1} - \mE_{\mf 2})}\ketbra{3}{1}\Big] \otimes \sum_{\eta_2, \eta_3}\ketbra{\eta_2 + 1, \eta_3}{\eta_2, \eta_3} \notag\\
    - s e^{-i \tau \mf \mE_{\mf 3}} \ketbra{1}{2} \otimes \sum_{\eta_2, \eta_3}\ketbra{\eta_2, \eta_3 - 1}{\eta_2, \eta_3}
    - s e^{i \tau \mf \mE_{\mf 3}} \ketbra{2}{1} \otimes \sum_{\eta_2, \eta_3}\ketbra{\eta_2, \eta_3 + 1}{\eta_2, \eta_3} \notag\\
    - s e^{i \tau \mf (\mE_{\mf 2} - \mE_{\mf 3})} \ketbra{2}{1} \otimes \sum_{\eta_2, \eta_3}\ketbra{\eta_2 + 1, \eta_3 - 1}{\eta_2, \eta_3}
    - s e^{-i \tau \mf (\mE_{\mf 2} - \mE_{\mf 3})} \ketbra{1}{2} \otimes \sum_{\eta_2, \eta_3}\ketbra{\eta_2 - 1, \eta_3 + 1}{\eta_2, \eta_3} \notag\\
    + \frac{\mf}{2} \Big(\mE_{\mf 2}\ketbra{1}{1} + \mE_{\mf 1} \ketbra{3}{3}\Big) \otimes \sum_{\eta_2, \eta_3}\ketbra{\eta_2, \eta_3}{\eta_2, \eta_3}\;. \label{eq:fl_ham_diamond_2D}
\end{align}
}
\noindent Figure~\ref{fig:2d_diamond} shows this hopping network on the 2D diamond lattice, that has the same spectrum as the 3D kagome Hamiltonian discussed in the main text.

The 1D diamond Floquet Hamiltonian can be extracted from the above Hamiltonian by eliminating the contributions along \(\eta_3\) coordinate.
{\small
\begin{align}
    \mh_{1D}(\tau)
    = -\Big[t \ketbra{1}{2} + t \ketbra{2}{1} + \big\{e^{i \varphi} + e^{-i \varphi + i \tau \mf \mE_{\mf 1}}\big\}\ketbra{2}{3} + \big\{e^{-i \varphi} + e^{i \varphi - i \tau \mf \mE_{\mf 1}}\big\} \ketbra{3}{2} + \ketbra{1}{3} + \ketbra{3}{1}\Big] \otimes \sum_{\eta_2}\ketbra{\eta_2}{\eta_2} \notag\\
    - \Big[t e^{-i \tau \mf \mE_{\mf 2}} \ketbra{1}{2} - e^{i \tau \mf(\mE_{\mf 1} - \mE_{\mf 2})}\ketbra{1}{3}\Big] \otimes \sum_{\eta_2}\ketbra{\eta_2 - 1}{\eta_2}
    - \Big[t e^{i \tau \mf \mE_{\mf 2}} \ketbra{2}{1} - e^{-i \tau \mf(\mE_{\mf 1} - \mE_{\mf 2})}\ketbra{3}{1}\Big] \otimes \sum_{\eta_2}\ketbra{\eta_2 + 1}{\eta_2} \notag\\
    + \frac{\mf}{2} \Big(\mE_{\mf 2}\ketbra{1}{1} + \mE_{\mf 1} \ketbra{3}{3}\Big) \otimes \sum_{\eta_2}\ketbra{\eta_2}{\eta_2}\;. \label{eq:fl_ham_diamond_1D}
\end{align}
}

\bibliography{general,flatband,local}

\end{document}